\PassOptionsToPackage{unicode}{hyperref}
\PassOptionsToPackage{hyphens}{url}
\PassOptionsToPackage{dvipsnames,svgnames*,x11names*}{xcolor}
\documentclass[
]{article}
\usepackage{lmodern}
\usepackage{amssymb,amsmath}
\usepackage{ifxetex,ifluatex}
\ifnum 0\ifxetex 1\fi\ifluatex 1\fi=0 % if pdftex
  \usepackage[T1]{fontenc}
  \usepackage[utf8]{inputenc}
  \usepackage{textcomp} % provide euro and other symbols
\else % if luatex or xetex
  \usepackage{unicode-math}
  \defaultfontfeatures{Scale=MatchLowercase}
  \defaultfontfeatures[\rmfamily]{Ligatures=TeX,Scale=1}
\fi
\IfFileExists{upquote.sty}{\usepackage{upquote}}{}
\IfFileExists{microtype.sty}{% use microtype if available
  \usepackage[]{microtype}
  \UseMicrotypeSet[protrusion]{basicmath} % disable protrusion for tt fonts
}{}
\makeatletter
\@ifundefined{KOMAClassName}{% if non-KOMA class
  \IfFileExists{parskip.sty}{%
    \usepackage{parskip}
  }{% else
    \setlength{\parindent}{0pt}
    \setlength{\parskip}{6pt plus 2pt minus 1pt}}
}{% if KOMA class
  \KOMAoptions{parskip=half}}
\makeatother
\usepackage{xcolor}
\IfFileExists{xurl.sty}{\usepackage{xurl}}{} % add URL line breaks if available
\IfFileExists{bookmark.sty}{\usepackage{bookmark}}{\usepackage{hyperref}}
\hypersetup{
  pdftitle={Quantitative View of the Structure of Institutional Scientific Collaborations Using the Examples of Halle, Jena and Leipzig},
  pdfauthor={Aliakbar Akbaritabar},
  colorlinks=true,
  linkcolor=Maroon,
  filecolor=Maroon,
  citecolor=Blue,
  urlcolor=blue,
  pdfcreator={LaTeX via pandoc}}
\usepackage[margin=1in]{geometry}
\usepackage{longtable,booktabs}
\usepackage{etoolbox}
\makeatletter
\patchcmd\longtable{\par}{\if@noskipsec\mbox{}\fi\par}{}{}
\makeatother
\IfFileExists{footnotehyper.sty}{\usepackage{footnotehyper}}{\usepackage{footnote}}
\makesavenoteenv{longtable}
\usepackage{graphicx}
\makeatletter
\def\maxwidth{\ifdim\Gin@nat@width>\linewidth\linewidth\else\Gin@nat@width\fi}
\def\maxheight{\ifdim\Gin@nat@height>\textheight\textheight\else\Gin@nat@height\fi}
\makeatother
\setkeys{Gin}{width=\maxwidth,height=\maxheight,keepaspectratio}
\makeatletter
\def\fps@figure{htbp}
\makeatother
\usepackage{pdflscape}
\usepackage{booktabs}
\usepackage{longtable}
\usepackage{array}
\usepackage{multirow}
\usepackage{wrapfig}
\usepackage{float}
\usepackage{colortbl}
\usepackage{pdflscape}
\usepackage{tabu}
\usepackage{threeparttable}
\usepackage{threeparttablex}
\usepackage[normalem]{ulem}
\usepackage{makecell}
\usepackage{xcolor}
\ifluatex
  \usepackage{selnolig}  % disable illegal ligatures
\fi
\newlength{\cslhangindent}
\newlength{\csllabelwidth}
\newenvironment{CSLReferences}[3] % #1 hanging-ident, #2 entry sp
 {% don't indent paragraphs
  \setlength{\parindent}{0pt}
  \ifodd #1 \everypar{\setlength{\hangindent}{\cslhangindent}}\ignorespaces\fi
  \ifnum #2 > 0
  \setlength{\parskip}{#3\baselineskip}
  \fi
 }%
 {}
\usepackage{calc} % for \widthof, \maxof

\title{Quantitative View of the Structure of Institutional Scientific Collaborations Using the Examples of Halle, Jena and Leipzig}
\author{Aliakbar Akbaritabar\footnote{German Centre for Higher Education Research and Science Studies (DZHW), Berlin, Germany; \href{mailto:akbaritabar@dzhw.eu}{\nolinkurl{akbaritabar@dzhw.eu}}; \href{mailto:akbaritabar@gmail.com}{\nolinkurl{akbaritabar@gmail.com}}; ORCID = 0000-0003-3828-1533 (Corresponding Author)}}
\date{}

\begin{document}
\maketitle
\begin{abstract}
Examining effectiveness of institutional scientific coalitions can inform future policies. This is a study on the structure of scientific collaborations in three cities in central Germany. Since 1995, the three universities of this region have formed and maintained a coalition which led to the establishment of an interdisciplinary center in 2012, i.e., German Center for Integrative Biodiversity Research (iDiv). We investigate whether the impact of the former coalition is evident in the region's structure of scientific collaborations and the scientific output of the new center. Using publications data from 1996-2018, we build co-authorship networks and identify the most cohesive communities in terms of collaboration, and compare them with communities identified based on publications presented as the scientific outcome of the coalition and new center on their website. Our results show that despite the highly cohesive structure of collaborations presented on the coalition website, there is still much potential to be realized. The newly established center has bridged the member institutions but not to a particularly strong level. We see that geographical proximity, collaboration policies, funding, and organizational structure alone do not ensure prosperous scientific collaboration structures. When new center's scientific output is compared with its regional context, observed trends become less conspicuous. Nevertheless, the level of success the coalition achieved could inform policy makers regarding other regions' scientific development plans.
\end{abstract}

\textbf{keywords}: Internationalization, Co-authorship Network Analysis, Bipartite Community Detection, Universitätsbund Halle Jena Leipzig, German Center for Integrative Biodiversity Research (iDiv)

\hypertarget{intro}{%
\section{Introduction}\label{intro}}

It is argued that scientific and complex economic activities are concentrated in urban and metropolitan areas (\protect\hyperlink{ref-ballandComplexEconomicActivities2020}{Balland et al., 2020}). This concentration might be the result of a selective process of spatial proximity between knowledge-producing and -demanding institutions (\protect\hyperlink{ref-rammerKnowledgeProximityFirm2020}{Rammer, Kinne, \& Blind, 2020}). In specific cases and within a densely populated metropolitan region, there might be remnants of a historical divide putting distance between institutions (\protect\hyperlink{ref-abbasiharoftehStillShadowWall2020}{Abbasiharofteh \& Broekel, 2020}).

This concentration of scientific activities in metropolitan and urban areas could inspire strategic coalitions between scientific institutions (e.g., academic and non-academic organizations). \protect\hyperlink{ref-akbaritabarBerlinQuantitativeView2020}{Akbaritabar} (\protect\hyperlink{ref-akbaritabarBerlinQuantitativeView2020}{2020}) showed that \emph{geographical proximity} and being located in a densely populated metropolitan area, i.e., Berlin, was not enough for institutional collaboration ties to form. A history of division and competition on the one hand and coalitions and policies to foster regional cooperation on the other hand can lead to a complex scientific landscape and structure of scientific collaborations.

Specific disciplines might present a tendency toward higher or lower internationalization of scientific collaborations. They might have a quality of being more or less interdisciplinary which can change over time (\protect\hyperlink{ref-cravenEvolutionInterdisciplinarityBiodiversity2019}{Craven et al., 2019}). These disciplines might have a global division of labor in that some researchers located in specific geographical areas might focus on specific themes or carry out only parts of the research work (e.g., data gathering and field work (\protect\hyperlink{ref-boshoffNeocolonialismResearchCollaboration2009}{Boshoff, 2009}) and hence the reason some researchers call for a more equal footing and benefit from scientific collaborations (\protect\hyperlink{ref-habelMoreEqualFooting2014}{Habel et al., 2014})). Biodiversity is one of these disciplines and previous research has shown a high degree of geographical specialization in this discipline. However, there might be an imbalance between the human resources and funding sources on the one hand versus where the biodiversity of natural environment exists on the other hand, leading to researchers from Europe and North America being the main producers of the knowledge about other geographical areas (mainly located in the global south) (\protect\hyperlink{ref-tydecksSpatialTopicalImbalances2018}{Tydecks, Jeschke, Wolf, Singer, \& Tockner, 2018}).

Studying structure of scientific collaborations in different contexts can reveal potential reasons behind observed strategic coalitions. There might be disciplinary, national, regional or continental specificities playing a role in determining which institutions collaborate. \protect\hyperlink{ref-shrumStructuresScientificCollaboration2007}{Shrum, Genuth, Carlson, Chompalov, \& Bijker} (\protect\hyperlink{ref-shrumStructuresScientificCollaboration2007}{2007}) studied \emph{structures of scientific collaborations} in multi-organizational research projects. They analyzed transcripts of interviews done in 1990s with scientists and administrative staff of different seniority levels from 53 projects in physical sciences. Although their sample included mainly large scale research projects and mostly from North America, however, they provided a useful typology of these collaborations. They used \emph{bureaucracy}\footnote{A formalized hierarchical structure with defined division of labor, goals and means to achieve them.} and \emph{technology}\footnote{Scientific endeavor in the studied fields is highly dependent on complex technology to gather information and produce scientific results.} as two main concepts to study how these collaborations form, raise funding, gather the data, own, share and analyze it and how they publish the results of the analysis. They investigated the \emph{division of labor}, \emph{leadership structure}, \emph{formalization} and \emph{hierarchy} in these collaborations. They found that formation process is mainly shaped by the level of complexity required by the collaboration. They proposed a fourfold typology based on organization of collaborations: bureaucratic (with a formalized structure and clear regulation of the decision making process which is mostly the case for larger collaborations), leaderless (similar to bureaucratic but with multiple leaders), non-specialized (semi-bureaucratic with lower levels of formalization) and participatory (a communitarian and democratic structure of decision making and collaboration). Even bureaucratic and semi-bureaucratic collaborations limited the scope of formal procedures to give members independence and autonomy in matters concerning production of scientific results. They presented an impersonal type of trust among scientific collaboration partners that is proved effective in these large scale projects where collaboration could happen between partners from diverse sectors who did not know each other beforehand (e.g., brokered collaboration) or partners with previous history of competition that might join forces to build a new technology. In these contexts, it is hard for individual researchers to interact and form interpersonal trust. Instead, bureaucracy plays an intermediating role to make these collaborations possible. Thus, researchers and governing bodies of these collaborations establish more (or in some cases less) formalized regulations on how to use the technology, share the ownership of the new instruments and obtained results and how to resolve conflicts. They presented a main exception, i.e., \emph{particle physics}, where researchers were being socialized from when they were postgraduate students to have a collaborative spirit. The main reason was the scope and complexity of the projects in this subfield that would be impossible without external collaborations among multiple organizations. They suspected whether studies on this exceptional subfield might have shaped the narrative around the prevalence of the \emph{team science}.

Subsequent to \protect\hyperlink{ref-akbaritabarBerlinQuantitativeView2020}{Akbaritabar} (\protect\hyperlink{ref-akbaritabarBerlinQuantitativeView2020}{2020})'s study on structure of scientific collaborations in Berlin metropolitan region, here, we focus on a geographical area in central Germany consisting of the three cities of Halle (Saale), Jena, and Leipzig (i.e., HJL cities). These cities have a total population of 843,790 (about 25\% of Berlin)\footnote{Population counts from: \url{https://worldpopulationreview.com/countries/cities/germany} accessed on October 8\textsuperscript{th} 2020.}. We selected these three cities as, since 1995, the three universities located there have maintained an agreement to foster interdisciplinary collaborations and formed a coalition named \emph{Unibund}\footnote{Universitätsbund Halle Jena Leipzig, \url{https://mitteldeutscher-unibund.de/ueber-uns/} accessed on October 8\textsuperscript{th} 2020.}. In addition, from 2012 this coalition in collaboration with eight other non-university research institutions has established an interdisciplinary center, i.e., German Center for Integrative Biodiversity Research (iDiv)\footnote{Deutsches Zentrum für integrative Biodiversitätsforschung (iDiv) Halle-Jena-Leipzig, \url{https://www.idiv.de/en/about-idiv.html} accessed on October 8\textsuperscript{th} 2020.} (see Table \ref{memberstable} for the list of members). iDiv received two funding phases from the German Science Foundation (DFG)\footnote{Deutsche Forschungsgemeinschaft, DFG.} for 2012 till 2016 and 2016 till 2020.

\begin{figure}

{\centering \includegraphics[width=0.7\linewidth,]{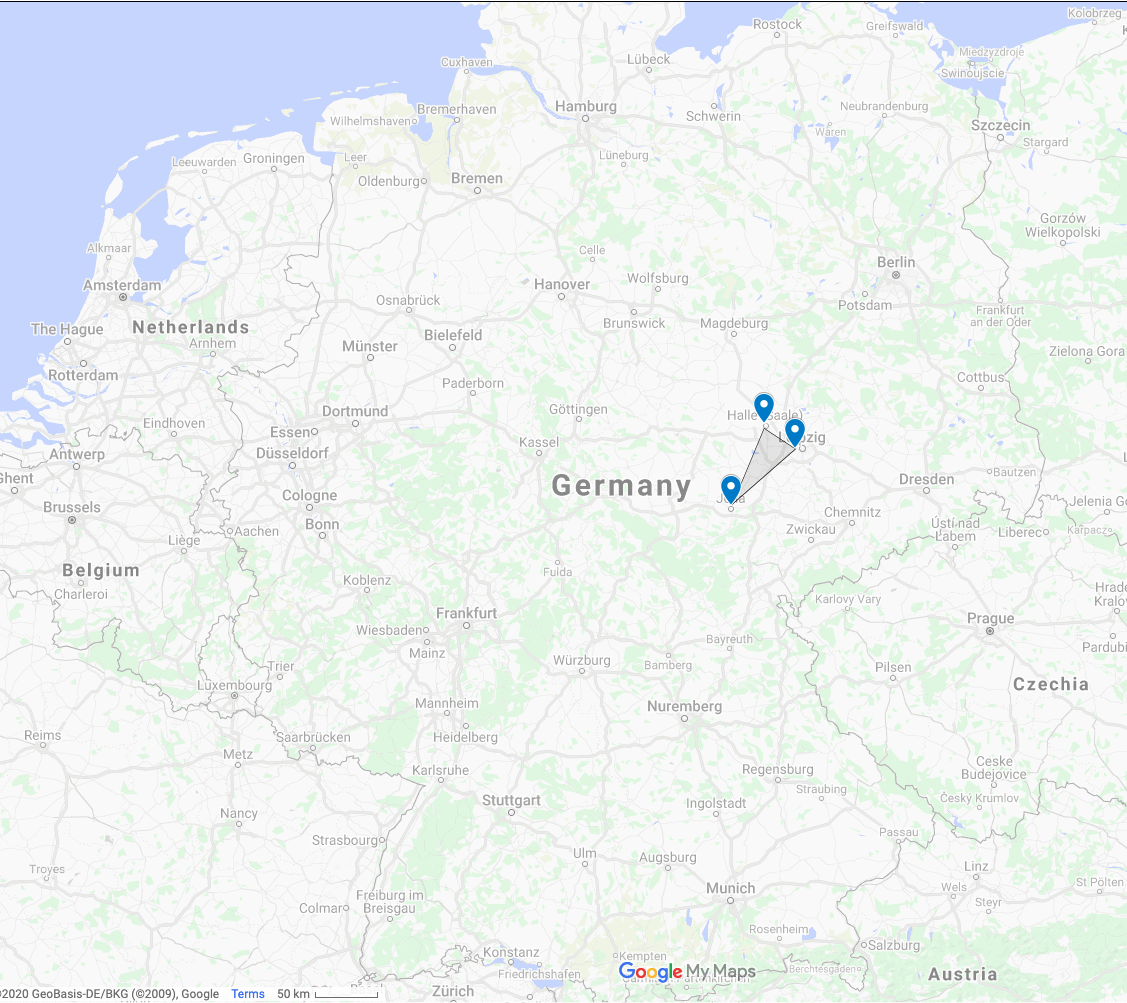} 

}

\caption{Approximate location of HJL cities and region in central Germany}\label{fig:map-of-hjl-cities-in-DEU}
\end{figure}

We analyze this strategic coalition as a natural experiment that has lasted 25 years and created a new organizational form. We aimed to investigate the coalition's effect in shaping the structure of scientific collaborations in this region and among its members.

Furthermore, we compare the structure of collaborations as represented by the coalition members on their website with the structure that we constructed based on the scientific output of the region. This gives us a baseline of comparison to determine the coalition's scientific performance versus its regional context. This is an interesting region as it does not comply with the idea of a centralized and concentrated metropolitan area (see Figure \ref{fig:map-of-hjl-cities-in-DEU}) such as Berlin (\protect\hyperlink{ref-akbaritabarBerlinQuantitativeView2020}{Akbaritabar} (\protect\hyperlink{ref-akbaritabarBerlinQuantitativeView2020}{2020}); \protect\hyperlink{ref-abbasiharoftehStillShadowWall2020}{Abbasiharofteh \& Broekel} (\protect\hyperlink{ref-abbasiharoftehStillShadowWall2020}{2020})) or other major cities (\protect\hyperlink{ref-ballandComplexEconomicActivities2020}{Balland et al.} (\protect\hyperlink{ref-ballandComplexEconomicActivities2020}{2020})). We explore how this top-down coalition is positioned in the scientific landscape of the region and structure of institutional collaborations.

Table \ref{memberstable} presents the list of Unibund and iDiv members. There were three universities (i.e., Martin Luther University Halle-Wittenberg, Friedrich Schiller University Jena, and Leipzig University) that from July 1995 formed Unibund and in collaboration with Helmholtz Centre for Environmental Research (UFZ), they established iDiv in 2012 and seven other institutions joined them in this interdisciplinary network.

\begin{longtable}[]{@{}lcc@{}}
\caption{Membership of institutions in Unibund and/or iDiv (Y = yes). \label{memberstable}}\tabularnewline
\toprule
\begin{minipage}[b]{0.26\columnwidth}\raggedright
Institution\strut
\end{minipage} & \begin{minipage}[b]{0.34\columnwidth}\centering
Unibund member?\strut
\end{minipage} & \begin{minipage}[b]{0.30\columnwidth}\centering
iDiv Member?\strut
\end{minipage}\tabularnewline
\midrule
\endfirsthead
\toprule
\begin{minipage}[b]{0.26\columnwidth}\raggedright
Institution\strut
\end{minipage} & \begin{minipage}[b]{0.34\columnwidth}\centering
Unibund member?\strut
\end{minipage} & \begin{minipage}[b]{0.30\columnwidth}\centering
iDiv Member?\strut
\end{minipage}\tabularnewline
\midrule
\endhead
\begin{minipage}[t]{0.26\columnwidth}\raggedright
1. Martin Luther University Halle-Wittenberg\footnote{\url{https://ror.org/05gqaka33}}\strut
\end{minipage} & \begin{minipage}[t]{0.34\columnwidth}\centering
Y\strut
\end{minipage} & \begin{minipage}[t]{0.30\columnwidth}\centering
Y\strut
\end{minipage}\tabularnewline
\begin{minipage}[t]{0.26\columnwidth}\raggedright
2. Friedrich Schiller University Jena\footnote{\url{https://ror.org/05qpz1x62}}\strut
\end{minipage} & \begin{minipage}[t]{0.34\columnwidth}\centering
Y\strut
\end{minipage} & \begin{minipage}[t]{0.30\columnwidth}\centering
Y\strut
\end{minipage}\tabularnewline
\begin{minipage}[t]{0.26\columnwidth}\raggedright
3. Leipzig University\footnote{\url{https://ror.org/03s7gtk40}}\strut
\end{minipage} & \begin{minipage}[t]{0.34\columnwidth}\centering
Y\strut
\end{minipage} & \begin{minipage}[t]{0.30\columnwidth}\centering
Y\strut
\end{minipage}\tabularnewline
\begin{minipage}[t]{0.26\columnwidth}\raggedright
4. Helmholtz Centre for Environmental Research -- UFZ\footnote{\url{https://ror.org/000h6jb29}}\strut
\end{minipage} & \begin{minipage}[t]{0.34\columnwidth}\centering
\strut
\end{minipage} & \begin{minipage}[t]{0.30\columnwidth}\centering
Y\strut
\end{minipage}\tabularnewline
\begin{minipage}[t]{0.26\columnwidth}\raggedright
5. Max Planck Institute for Biogeochemistry (MPI BGC)\footnote{\url{https://ror.org/051yxp643}}\strut
\end{minipage} & \begin{minipage}[t]{0.34\columnwidth}\centering
\strut
\end{minipage} & \begin{minipage}[t]{0.30\columnwidth}\centering
Y\strut
\end{minipage}\tabularnewline
\begin{minipage}[t]{0.26\columnwidth}\raggedright
6. Max Planck Institute for Chemical Ecology (MPI CE)\footnote{\url{https://ror.org/02ks53214}}\strut
\end{minipage} & \begin{minipage}[t]{0.34\columnwidth}\centering
\strut
\end{minipage} & \begin{minipage}[t]{0.30\columnwidth}\centering
Y\strut
\end{minipage}\tabularnewline
\begin{minipage}[t]{0.26\columnwidth}\raggedright
7. Max Planck Institute for Evolutionary Anthropology\footnote{\url{https://ror.org/02a33b393}}\strut
\end{minipage} & \begin{minipage}[t]{0.34\columnwidth}\centering
\strut
\end{minipage} & \begin{minipage}[t]{0.30\columnwidth}\centering
Y\strut
\end{minipage}\tabularnewline
\begin{minipage}[t]{0.26\columnwidth}\raggedright
8. Leibniz Institute German Collection of Microorganisms and Cell Cultures (DSMZ)\footnote{\url{https://ror.org/02tyer376}}\strut
\end{minipage} & \begin{minipage}[t]{0.34\columnwidth}\centering
\strut
\end{minipage} & \begin{minipage}[t]{0.30\columnwidth}\centering
Y\strut
\end{minipage}\tabularnewline
\begin{minipage}[t]{0.26\columnwidth}\raggedright
9. Leibniz Institute of Plant Biochemistry (IPB)\footnote{\url{https://ror.org/01mzk5576}}\strut
\end{minipage} & \begin{minipage}[t]{0.34\columnwidth}\centering
\strut
\end{minipage} & \begin{minipage}[t]{0.30\columnwidth}\centering
Y\strut
\end{minipage}\tabularnewline
\begin{minipage}[t]{0.26\columnwidth}\raggedright
10. Leibniz Institute of Plant Genetics and Crop Plant Research (IPK)\footnote{\url{https://ror.org/02skbsp27}}\strut
\end{minipage} & \begin{minipage}[t]{0.34\columnwidth}\centering
\strut
\end{minipage} & \begin{minipage}[t]{0.30\columnwidth}\centering
Y\strut
\end{minipage}\tabularnewline
\begin{minipage}[t]{0.26\columnwidth}\raggedright
11. Leibniz Institute Senckenberg Museum of Natural History (SMNG)\footnote{\url{https://ror.org/05jv9s411}}\strut
\end{minipage} & \begin{minipage}[t]{0.34\columnwidth}\centering
\strut
\end{minipage} & \begin{minipage}[t]{0.30\columnwidth}\centering
Y\strut
\end{minipage}\tabularnewline
\bottomrule
\end{longtable}

iDiv is the youngest among seven research centers currently funded by the DFG. The goal of the DFG's first funding phase (2012-2016) was to establish the center to facilitate cooperation in later stages e.g., in the second funding phase (2016-2020). This is the first time the DFG has funded a research center developed by a consortium of three universities and eight non-university research institutions, particularly as the institutions are located in three different federal states. iDiv is a disciplinary integration of biology, chemistry, physics, geosciences, economics, social sciences and computer sciences. The biodiversity field is known to be highly interdisciplinary (\protect\hyperlink{ref-tydecksSpatialTopicalImbalances2018}{Tydecks, Jeschke, Wolf, Singer, \& Tockner, 2018}; \protect\hyperlink{ref-cravenEvolutionInterdisciplinarityBiodiversity2019}{Craven et al., 2019}), nevertheless, the main focus of iDiv is in natural sciences. iDiv has four main research areas: 1) biodiversity patterns, 2) biodiversity processes, 3) biodiversity functions, and 4) biodiversity and society. iDiv has successfully obtained support from the local government bodies. From 2012 Saxony (Freistaat Sachsen) provided 2,600 m\textsuperscript{2} space for the buildings and offices in Leipziger BioCity. Thuringia and Saxony-Anhalt (Thüringen und Sachsen-Anhalt) have four professorship chairs for this cooperation. iDiv's website states having 350 employees from 30 different nations.

Based on this introduction and background, we explore the interplay between different contextual variables and geographical proximity to investigate the structure of scientific collaborations in the HJL region. We specifically focus on the Unibund and iDiv to identify how these strategic coalitions are positioned in the structure of the region's scientific collaborations. Therefore, after presenting the effect of \emph{disambiguation} of scientific organizations' names on the structure of scientific collaborations as our main methodological contribution, we formulate and investigate the following macro, quantitative, and exploratory research questions:

\begin{itemize}
\item
  \textbf{RQ1}: How \emph{collaborative} and \emph{internationalized} is the scientific landscape of the HJL region?
\item
  \textbf{RQ2}: Are there \emph{disciplinary} differences in the rate of collaborative and internationalized scientific work?
\item
  \textbf{RQ3}: How regionally- or continentally-oriented is scientific collaboration in the HJL region?
\item
  \textbf{RQ4}: How sector-oriented is scientific collaboration in the HJL region?
\item
  \textbf{RQ5}: Are there specific disciplinary, sectoral, national or continental \emph{cohesive subgroups} driving the scientific collaborations in the HJL region?
\item
  \textbf{RQ6}: Can we find evidence of Unibund's effect as a strategic coalition on the structure of scientific collaborations network in the HJL region?
\item
  \textbf{RQ7}: Can we find evidence of iDiv's effect as a new organizational form on the structure of scientific collaborations network in the HJL region?
\item
  \textbf{RQ8}: How cohesively do Unibund and iDiv member institutions collaborate among themselves?
\end{itemize}

The contributions of this paper are threefold: 1) We focus on the scientific output of the HJL region, present the share of collaborative works, and identify the share of international collaborations. We differentiate HJL region, Berlin, Germany, Europe and continental regions worldwide to investigate possible groupings. We examine the effectiveness of the institutional scientific coalition in HJL region. To do so, we compare Unibund and iDiv's self-represented scientific output and institutional collaboration structure with the structure constructed from the larger context of HJL region. 2) We compare all OECD scientific fields on the structure of collaborations and we include a sectoral view based on the type of organization involved in the collaborations. 3) We present the effect of the organization name disambiguation on results and employ a bipartite network modeling and bipartite community detection approach and present how it can be useful in the identification of denser collaborative structures.

We present our data source and modeling strategy in the \protect\hyperlink{datamethods}{Data and Methods} section, and summarize our main findings in the \protect\hyperlink{results}{Results} section, followed by the \protect\hyperlink{conclusions}{Discussion and Conclusions} section.

\hypertarget{datamethods}{%
\section{Data and Methods}\label{datamethods}}

We used two publication datasets. First, to delineate all scientific output of the iDiv, we used all publications listed on their website\footnote{\url{https://www.idiv.de/en/research/publications.html} (accessed on July 30\textsuperscript{th} 2020)}. We found a total of 2,583 records (including 18 code, 66 data, 2,393 DOI, and 106 PDF). We excluded those without or with problematic DOIs and searched the remaining records (2,371 unique DOIs) in Scopus 2019 from the German Bibliometrics Competence Center (KB)\footnote{Kompetenzzentrum Bibliometrie (KB), \url{http://bibliometrie.info}}. The KB database was updated until April 2019, thus it did not include all publications from 2019 and 2020. Therefore, we extracted matching records up to the end of 2018 (1,749 unique records, 74\%).

In parallel, to use as a baseline for comparison, we extracted all \emph{article}, \emph{review} and \emph{conference proceedings} documents published from the beginning of the database in 1996 until the end of 2018 with at least one authoring organization from \emph{Halle}, \emph{Saale}, \emph{Jena} or \emph{Leipzig}. Our purpose here was to find both publications from \emph{iDiv} and its predecessor, \emph{Unibund}, and compare how the former coalition and its established interdisciplinary center are positioned in the larger structure of scientific collaborations in the region. Please note that Unibud had a larger scope than iDiv's disciplinary focus in biodiversity, nevertheless, we investigate whether prior coalition fostered closer collaboration ties in case of the iDiv.

In both cases (i.e., iDiv web and HJL cities), we included organizations worldwide who have collaborated with at least one Unibund or iDiv member. Our level of analysis was \emph{scientific organizations} (i.e., each academic or non-academic affiliation mentioned in a publication) and we do not investigate lower levels e.g., authors, since our goal is to identify the structure of scientific collaborations among organizations.

The data included meta-data for each publication such as \emph{publication year}, \emph{title}, \emph{affiliation addresses}, \emph{scientific field}, \emph{journal name} and \emph{document type}. We used a mapping of publications to OECD scientific fields based on Scopus ASJC\footnote{All Science Journal Classification} that reduces the number of subject categories from 33 to 6. We compared the aggregate data of iDiv web and HJL cities with trends of different OECD scientific fields i.e., \emph{Agricultural Sciences} (AS), \emph{Engineering Technology} (ET), \emph{Natural Sciences} (NS), \emph{Medical and Health Sciences} (MHS), \emph{Humanities} (H) and \emph{Social Sciences} (SS). As some publications are assigned to multiple fields, in the aggregate analysis, we used the first assignment of each publication, but in a single field view, we take all publications with any class assignment in the given field, thus, interdisciplinary publications are covered separately in all their assigned fields.

We used the complete string of affiliation addresses to disambiguate institutions names and obtain further information (i.e., country, geographical coordinates (longitude and latitude) of the main address and type of organization as \emph{education}, \emph{non-profit}\footnote{Organization that uses its surplus revenue to achieve its goals. Includes charities and other non-government research funding bodies. Example, the Max Planck Society itself (\href{https://www.grid.ac/institutes/grid.4372.2}{grid.4372.2})}, \emph{company}, \emph{government}, \emph{health-care}, \emph{facility}\footnote{A building or facility dedicated to research of a specific area, usually contains specialized equipment. Includes telescopes, observatories and particle accelerators. Example: child institutes of the Max Planck Society (e.g., Max Planck Institute for Demographic Research, \href{https://www.grid.ac/institutes/grid.419511.9}{grid.419511.9})}, \emph{archive}\footnote{Repository of documents, artifacts, or specimens. Includes libraries and museums that are not part of a university. Example, New York Public Library (\href{https://www.grid.ac/institutes/grid.429888.7}{grid.429888.7})} and other, see Global Research Identifier Database (GRID) policies\footnote{\url{https://www.grid.ac/pages/policies}}) from the Research Organization Registry (ROR) API\footnote{\url{https://ror.org/about}}. ROR aggregates data from GRID, ISNI\footnote{International Standard Name Identifier, \url{https://isni.org/}}, Crossref and Wikidata\footnote{\url{https://www.wikidata.org}}. We used the ROR snapshot from November 7\textsuperscript{th} 2019. This disambiguation method takes into account different name spellings, including misspellings, acronyms, and multiple languages. It is important to note that in disambiguated versions of the data presented in the results section, we include only publications for which all contributing organizations are disambiguated and exclude those publications with one or more non-disambiguated co-authoring organizations. This reduces our coverage of publications, but gives more reliable results as non-disambiguated data can lead to multiple representation of similar organizations in the structure and produce misleading results.

To evaluate the quality of ROR disambiguation results, we chose a random sample of 100 organizations. A research assistant searched online for the organization names from Scopus and the disambiguated name from ROR to find the original address and confronted the two names to determine whether the match was reliable. The results indicated that in only 4\% of the cases a false match was assigned and 96\% of cases were reliable matches.

We constructed bipartite co-authorship networks (\protect\hyperlink{ref-breigerDualityPersonsGroups1974}{Breiger, 1974}) using ties between publications and \emph{organizations} (\protect\hyperlink{ref-katzWhatResearchCollaboration1997}{Katz \& Martin, 1997}). We treated each single publication as an event where organizations, through collaboration between their members, interacted to produce an academic text (\protect\hyperlink{ref-biancaniSocialNetworksResearch2013}{Biancani \& McFarland, 2013}). Studies on co-authorship networks usually use a one-mode projection of these bipartite networks (\protect\hyperlink{ref-newmanScientificCollaborationNetworks2001}{Newman, 2001b}, \protect\hyperlink{ref-newmanScientificCollaborationNetworks2001a}{2001a}). The problem with this projection is twofold. First, different structures in two-mode networks are projected to the same one-mode structure which causes information about the underlying structure to be lost. Second, the one-mode projection can present an artificially higher density and connectivity due to publications with high number of authors that project to maximally connected cliques. By adopting methods and modeling strategies specifically developed for \emph{bipartite networks} we are able to resolve the shortcomings.

To identify possible geographical, disciplinary and/or sector-based coalitions between scientific organizations, we extracted the largest connected component of the co-authorship network, i.e., the giant component, and investigated it further. Our aim was to see if there are cohesive subgroups of organizations preferentially collaborating \emph{among} themselves. We investigated the potential underlying factors behind these groupings.

To identify communities of co-authorship, we used \emph{bipartite community detection} by \emph{Constant Potts model} (CPM). CPM is a specific version of Potts model (\protect\hyperlink{ref-reichardtDetectingFuzzyCommunity2004}{Reichardt \& Bornholdt, 2004}) proposed by \protect\hyperlink{ref-traagNarrowScopeResolutionlimitfree2011}{Traag, Van Dooren, \& Nesterov} (\protect\hyperlink{ref-traagNarrowScopeResolutionlimitfree2011}{2011}) as a \emph{resolution-limit-free} method. It resolves the resolution limit problem in modularity (\protect\hyperlink{ref-newmanDetectingCommunityStructure2004}{Newman, 2004}) which can obstruct detection of small communities in large networks (\protect\hyperlink{ref-traagLouvainLeidenGuaranteeing2019}{Traag, Waltman, \& van Eck, 2019}). We use the implementation in the \emph{Leidenalg}\footnote{\url{https://github.com/vtraag/leidenalg}} library in Python. Community detection emphasizes the importance of links \emph{within} communities rather than those \emph{between} them. CPM uses a resolution parameter \(\gamma\) (i.e., ``\emph{constant}'' in the name), leading to communities such that the link density between the communities (external density) is lower than \(\gamma\) and the link density within communities (internal density) is more than \(\gamma\). We set \(6 \times 10^{-3}\) as the resolution parameter in the case of all networks, i.e., aggregate data and scientific fields. We chose this parameter after exploration of the number of communities detected in contrast to the number of organizations and publications included in each bipartite community to prevent having too many or too few organizations in communities or an extremely high number of communities which will not be interpretable.

\hypertarget{results}{%
\section{Results}\label{results}}

Table \ref{tab:description-different-disambiguation} presents descriptive information about the co-authorship networks constructed using non-disambiguated and disambiguated data from iDiv website and HJL cities. The lower number of connected components and decrease in the number of organizations in the disambiguated version of the networks show that in the non-disambiguated data there were multiple representations of the same institution that needed to be resolved before further analysis. Note that the disambiguation results differ slightly between the two cases and the overlap from the web sample to cities is 828 publications (90\%), which is due to non-matching DOIs. Note also that we excluded publications for which one or some of coauthoring organizations were not successfully disambiguated, which increases the quality and reliability of our data but decreases our coverage. Table \ref{tab:description-different-field-networks} shows the networks constructed for each scientific field with web and cities data after disambiguation, which shows the highest share of publications were in Natural Sciences (NS). Ninety-three percent of iDiv's publications were categorized as NS, which aligns with the center's primary focus in this field.

\begin{table}

\caption{\label{tab:description-different-disambiguation}HJL organizations co-authorship networks using non-disambiguated and disambiguated data (G = giant component, Scopus 1996-2018)}
\centering
\resizebox{\linewidth}{!}{
\begin{tabular}[t]{l|r|r|r|r}
\hline
\cellcolor{lightgray}{\textcolor{black}{\textbf{Metrics                              }}} & \cellcolor{lightgray}{\textcolor{black}{\textbf{  Non disambiguated web}}} & \cellcolor{lightgray}{\textcolor{black}{\textbf{  Disambiguated web}}} & \cellcolor{lightgray}{\textcolor{black}{\textbf{  Non disambiguated cities}}} & \cellcolor{lightgray}{\textcolor{black}{\textbf{  Disambiguated cities}}}\\
\hline
N. of connected components & 228 & 1 & 61,906 & 22\\
\hline
N. of biparitite nodes & 14,577 & 1,604 & 617,248 & 86,233\\
\hline
N. of biparitite edges & 16,485 & 4,312 & 631,291 & 168,078\\
\hline
\% of biparitite nodes in G & 87 & 100 & 55 & 100\\
\hline
\% of biparitite edges in G & 90 & 100 & 65 & 100\\
\hline
\cellcolor{lightgray}{\textcolor{black}{\textbf{N. of organizations}}} & \cellcolor{lightgray}{\textcolor{black}{\textbf{12,828}}} & \cellcolor{lightgray}{\textcolor{black}{\textbf{686}}} & \cellcolor{lightgray}{\textcolor{black}{\textbf{462,426}}} & \cellcolor{lightgray}{\textcolor{black}{\textbf{6,289}}}\\
\hline
N. of organizations in G & 11,243 & 686 & 264,938 & 6,268\\
\hline
\cellcolor{lightgray}{\textcolor{black}{\textbf{N. of publications}}} & \cellcolor{lightgray}{\textcolor{black}{\textbf{1,749}}} & \cellcolor{lightgray}{\textcolor{black}{\textbf{918}}} & \cellcolor{lightgray}{\textcolor{black}{\textbf{154,822}}} & \cellcolor{lightgray}{\textcolor{black}{\textbf{79,944}}}\\
\hline
N. of publications in G & 1,462 & 918 & 75,055 & 79,894\\
\hline
\end{tabular}}
\end{table}

\begin{table}

\caption{\label{tab:description-different-field-networks}HJL organizations co-authorship networks in different OECD scientific fields (w = web, c = cities, G = giant component)}
\centering
\resizebox{\linewidth}{!}{
\begin{tabular}[t]{l|r|r|r|r|r|r|r|r|r|r|r|r}
\hline
\cellcolor{lightgray}{\textcolor{black}{\textbf{Metrics                              }}} & \cellcolor{lightgray}{\textcolor{black}{\textbf{  AS\_w}}} & \cellcolor{lightgray}{\textcolor{black}{\textbf{  AS\_c}}} & \cellcolor{lightgray}{\textcolor{black}{\textbf{  ET\_w}}} & \cellcolor{lightgray}{\textcolor{black}{\textbf{  ET\_c}}} & \cellcolor{lightgray}{\textcolor{black}{\textbf{  H\_w}}} & \cellcolor{lightgray}{\textcolor{black}{\textbf{  H\_c}}} & \cellcolor{lightgray}{\textcolor{black}{\textbf{  MHS\_w}}} & \cellcolor{lightgray}{\textcolor{black}{\textbf{  MHS\_c}}} & \cellcolor{lightgray}{\textcolor{black}{\textbf{  NS\_w}}} & \cellcolor{lightgray}{\textcolor{black}{\textbf{   NS\_c}}} & \cellcolor{lightgray}{\textcolor{black}{\textbf{  SS\_w}}} & \cellcolor{lightgray}{\textcolor{black}{\textbf{  SS\_c}}}\\
\hline
N. of connected components & 4 & 8 & 1 & 18 & 1 & 18 & 4 & 17 & 1 & 17 & 1 & 19\\
\hline
N. of biparitite nodes & 505 & 7,881 & 68 & 19,511 & 6 & 2,993 & 316 & 31,317 & 1,511 & 60,682 & 144 & 9,842\\
\hline
N. of biparitite edges & 1,026 & 14,279 & 87 & 34,578 & 5 & 4,080 & 578 & 55,259 & 3,966 & 122,102 & 216 & 15,880\\
\hline
\% of biparitite nodes in G & 98 & 100 & 100 & 100 & 100 & 98 & 98 & 100 & 100 & 100 & 100 & 99\\
\hline
\% of biparitite edges in G & 99 & 100 & 100 & 100 & 100 & 99 & 99 & 100 & 100 & 100 & 100 & 100\\
\hline
\cellcolor{lightgray}{\textcolor{black}{\textbf{N. of organizations}}} & \cellcolor{lightgray}{\textcolor{black}{\textbf{279}}} & \cellcolor{lightgray}{\textcolor{black}{\textbf{1,816}}} & \cellcolor{lightgray}{\textcolor{black}{\textbf{51}}} & \cellcolor{lightgray}{\textcolor{black}{\textbf{2,469}}} & \cellcolor{lightgray}{\textcolor{black}{\textbf{5}}} & \cellcolor{lightgray}{\textcolor{black}{\textbf{687}}} & \cellcolor{lightgray}{\textcolor{black}{\textbf{167}}} & \cellcolor{lightgray}{\textcolor{black}{\textbf{3,357}}} & \cellcolor{lightgray}{\textcolor{black}{\textbf{655}}} & \cellcolor{lightgray}{\textcolor{black}{\textbf{5,316}}} & \cellcolor{lightgray}{\textcolor{black}{\textbf{95}}} & \cellcolor{lightgray}{\textcolor{black}{\textbf{1,646}}}\\
\hline
N. of organizations in G & 270 & 1,809 & 51 & 2,449 & 5 & 669 & 163 & 3,341 & 655 & 5,300 & 95 & 1,623\\
\hline
\cellcolor{lightgray}{\textcolor{black}{\textbf{N. of publications}}} & \cellcolor{lightgray}{\textcolor{black}{\textbf{226}}} & \cellcolor{lightgray}{\textcolor{black}{\textbf{6,065}}} & \cellcolor{lightgray}{\textcolor{black}{\textbf{17}}} & \cellcolor{lightgray}{\textcolor{black}{\textbf{17,042}}} & \cellcolor{lightgray}{\textcolor{black}{\textbf{1}}} & \cellcolor{lightgray}{\textcolor{black}{\textbf{2,306}}} & \cellcolor{lightgray}{\textcolor{black}{\textbf{149}}} & \cellcolor{lightgray}{\textcolor{black}{\textbf{27,960}}} & \cellcolor{lightgray}{\textcolor{black}{\textbf{856}}} & \cellcolor{lightgray}{\textcolor{black}{\textbf{55,366}}} & \cellcolor{lightgray}{\textcolor{black}{\textbf{49}}} & \cellcolor{lightgray}{\textcolor{black}{\textbf{8,196}}}\\
\hline
N. of publications in G & 223 & 6,055 & 17 & 17,008 & 1 & 2,279 & 146 & 27,935 & 856 & 55,337 & 49 & 8,158\\
\hline
\end{tabular}}
\end{table}

\begin{figure}

{\centering \includegraphics[width=1\linewidth,]{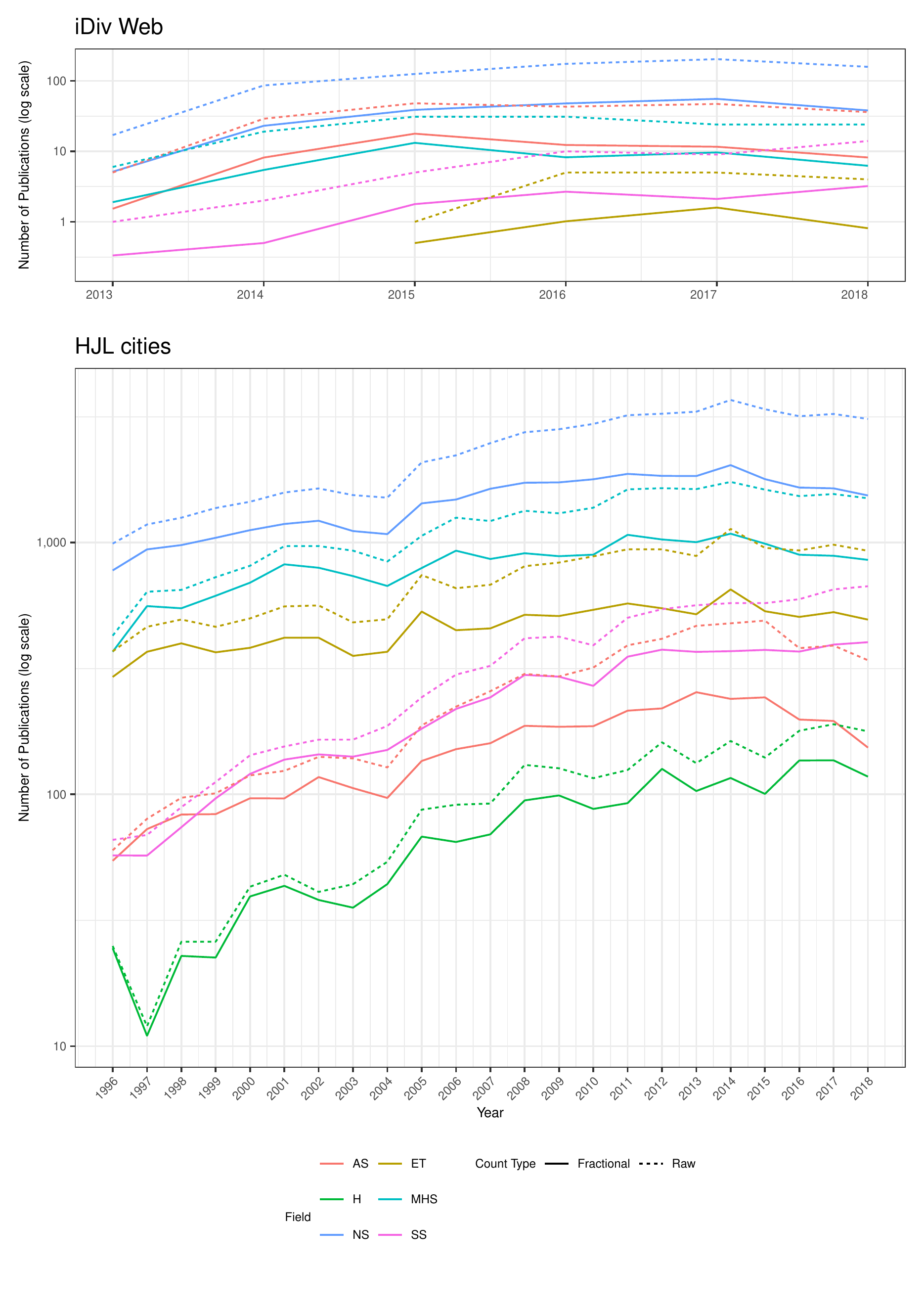} 

}

\caption{Raw and fractional count of HJL publications by OECD fields (top, web, bottom, cities, 1996-2018, Scopus, fractional count based on organizations)}\label{fig:range-of-publications}
\end{figure}

Figure \ref{fig:range-of-publications} presents a temporal comparison between scientific fields on the one hand and publications presented on iDiv website (top) and three cities (bottom) on the other hand. Humanities and Social Sciences (which are green and red lines at the bottom) have sharply increased, which is the case for all Germany previously presented in \protect\hyperlink{ref-stahlschmidtPerformanceStructuresGerman2019}{Stahlschmidt, Stephen, \& Hinze} (\protect\hyperlink{ref-stahlschmidtPerformanceStructuresGerman2019}{2019}) and has to do with both an actual increase in publications, and increase in Scopus's coverage. For most fields, the gap between the two lines with the same color (i.e., raw and fractional counts) are stable which shows the maturity of the collaboration trajectory in the HJL region (\textbf{RQ1-2}).

\begin{figure}

{\centering \includegraphics[width=1\linewidth,]{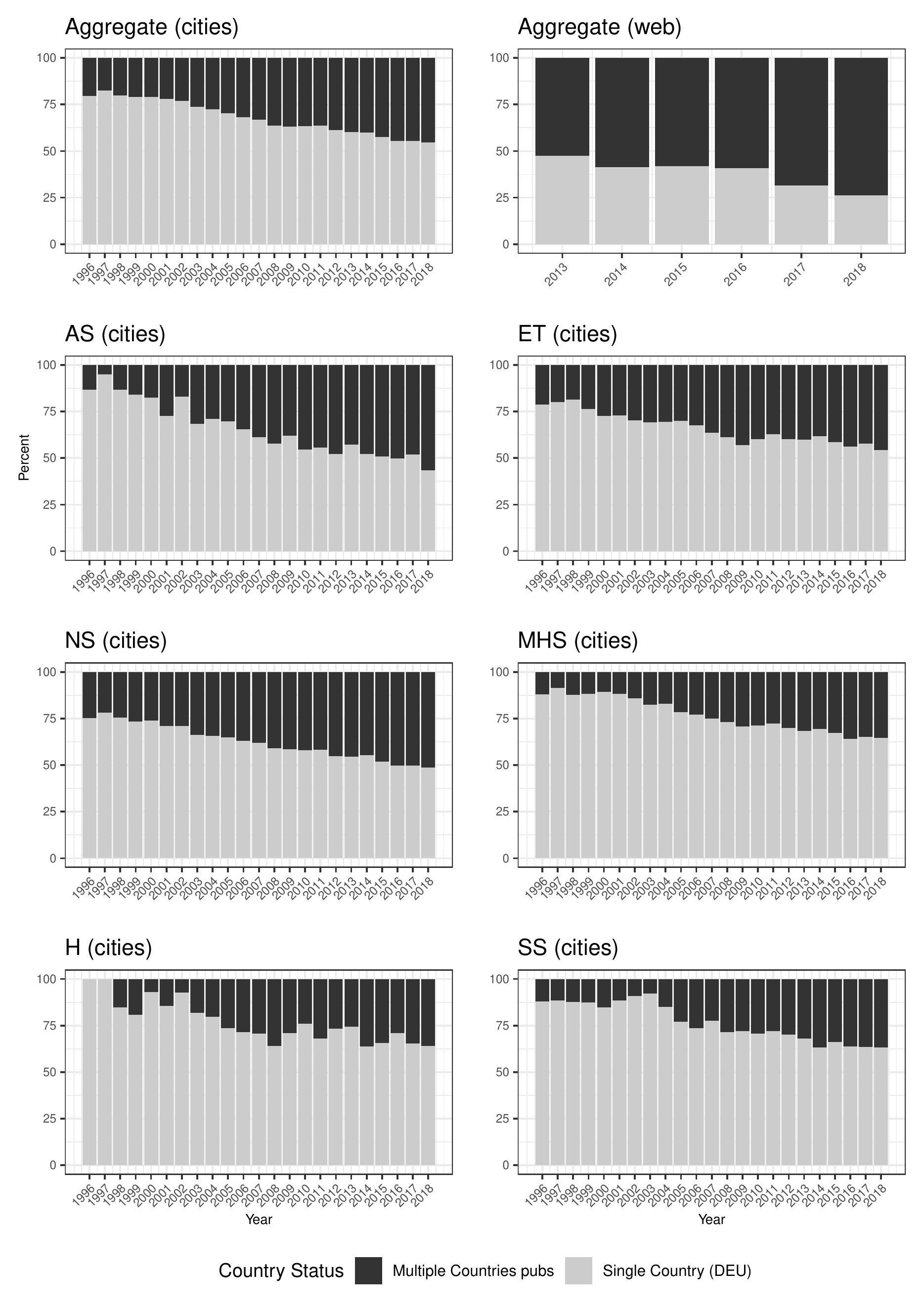} 

}

\caption{Share of intra-Germany versus multiple country co-authorship, top row, aggregate in cities and web, bottom rows, different fields for cities data (1996-2018, Scopus)}\label{fig:single-multiple-country-publications-fields-web-cities}
\end{figure}

Figure \ref{fig:single-multiple-country-publications-fields-web-cities} presents the level of internationalization of collaborations (\textbf{RQ1-2}). iDiv (shown in the uppermost part on the right) is a highly exceptional case in that from 2013 it starts with about 50\% international publications and arrives at about 75\% in 2018. The trends of the HJL cities show the prevalence of intra-Germany collaborations has decreased in the most recent years. Conversely, international collaboration has increased in all fields, with AS, NS, and to a lesser extent ET, holding the highest rates of internationalization, although these are still below 50\% in most years.

\begin{landscape}

\begin{figure}

{\centering \includegraphics[width=1\linewidth,]{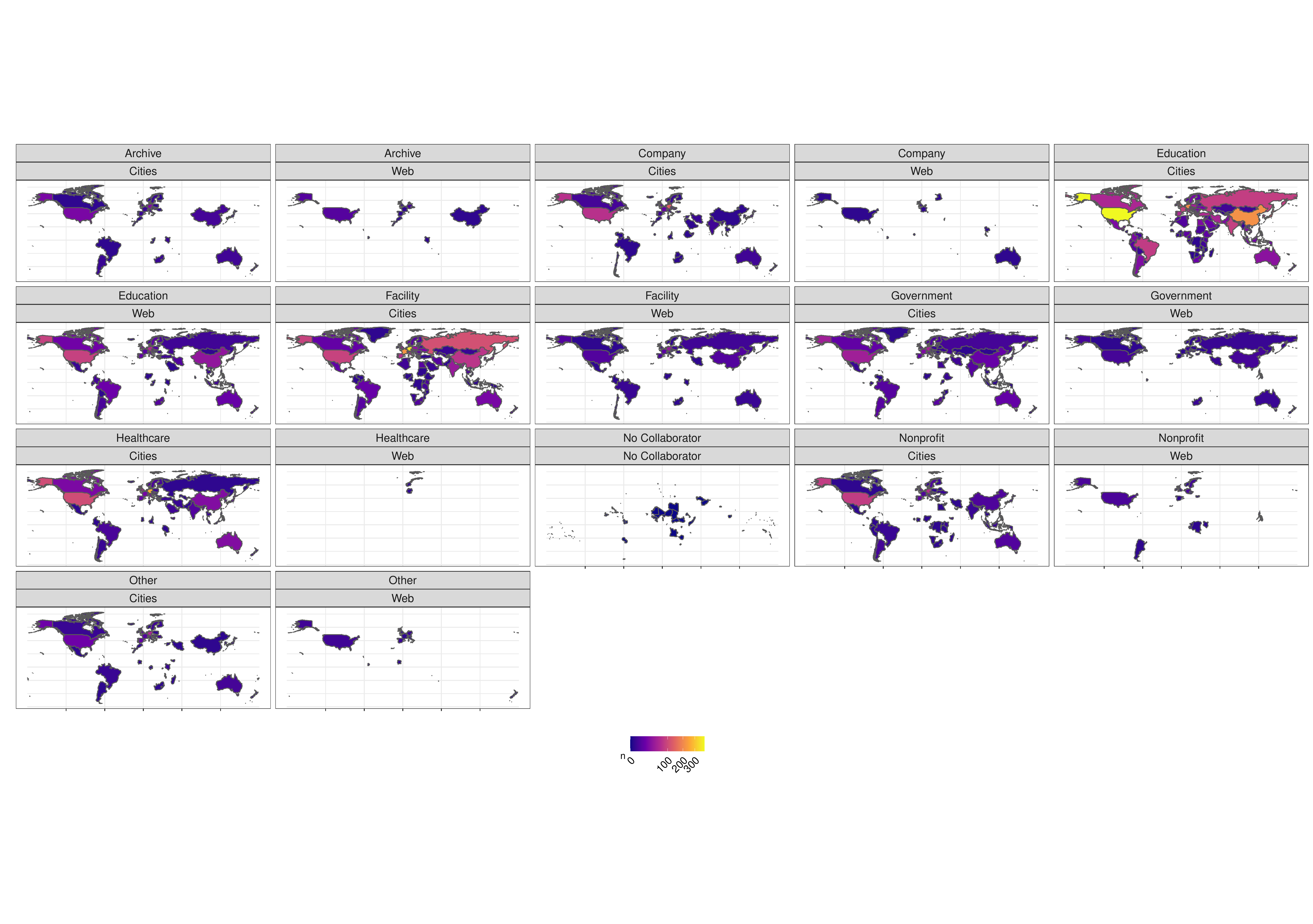} 

}

\caption{Countries worldwide collaborating with HJL (cities) or iDiv (web) by sector (color: N. of organizations. If a country does not have presence in a sector, it is shown in "No Collaborator")}\label{fig:map-of-orgs-type-world-web-cities}
\end{figure}

\end{landscape}

Figure \ref{fig:map-of-orgs-type-world-web-cities} presents a macro picture of the sectoral diversity (\textbf{RQ3-4}) of the institutions collaborating with HJL region (i.e., cities) or iDiv (i.e., web). In some countries (e.g., the more industrially developed and western ones indicated with warmer colors) there are representatives from all sectors active in research and publishing. Education and facility (e.g., Max Planck society institutes or Leibniz society institutes) were the most prolific sectors. There was an interesting difference between iDiv's collaborators worldwide and the picture from all institutions in the HJL region (including iDiv members). For example, while in education and facility, the trend of iDiv's collaborations was similar to HJL region, regarding companies (e.g., business sector) and health-care, it is highly different from the HJL region and shows the specific sectoral focus of this new organizational form. Table \ref{tab:description-different-org-types-web-cities} presents the top five countries in each sector, further highlighting the differences between the web and cities datasets and the dominance of education and facility organisations in research activity. The USA and China are the only collaborators among these five located outside Europe.

\begin{table}

\caption{\label{tab:description-different-org-types-web-cities}Five countries with the highest number of organizations by sector (GRID data based on HJL sample for web and cities 1996-2018)}
\centering
\fontsize{9}{11}\selectfont
\begin{tabular}[t]{l|l|r|r}
\hline
Organization sector & Country code & Count Web & Count Cities\\
\hline
 & DEU & 4 & 26\\

 & USA & 8 & 22\\

 & GBR & 2 & 7\\

 & CHN & 1 & 4\\

\multirow{-5}{*}{\raggedright\arraybackslash Archive} & ESP & 1 & 2\\
\cline{1-4}
 & DEU &  & 132\\

 & USA & 1 & 70\\

 & GBR & 2 & 13\\

 & ESP &  & 3\\

\multirow{-5}{*}{\raggedright\arraybackslash Company} & CHN &  & 1\\
\cline{1-4}
 & USA & 89 & 394\\

 & CHN & 35 & 213\\

 & DEU & 63 & 206\\

 & GBR & 28 & 111\\

\multirow{-5}{*}{\raggedright\arraybackslash Education} & ESP & 13 & 58\\
\cline{1-4}
 & DEU & 37 & 352\\

 & USA & 7 & 91\\

 & CHN & 7 & 74\\

 & ESP & 5 & 65\\

\multirow{-5}{*}{\raggedright\arraybackslash Facility} & GBR & 4 & 43\\
\cline{1-4}
 & DEU & 2 & 50\\

 & USA & 4 & 46\\

 & CHN & 4 & 14\\

 & GBR &  & 13\\

\multirow{-5}{*}{\raggedright\arraybackslash Government} & ESP & 1 & 11\\
\cline{1-4}
 & DEU & 1 & 263\\

 & USA &  & 105\\

 & GBR &  & 56\\

 & CHN &  & 26\\

\multirow{-5}{*}{\raggedright\arraybackslash Healthcare} & ESP &  & 15\\
\cline{1-4}
 & USA & 5 & 83\\

 & DEU & 4 & 68\\

 & GBR & 4 & 18\\

 & ESP & 1 & 11\\

\multirow{-5}{*}{\raggedright\arraybackslash Nonprofit} & CHN &  & 6\\
\cline{1-4}
 & DEU & 2 & 59\\

 & USA & 3 & 17\\

 & ESP & 2 & 13\\

 & GBR & 1 & 9\\

\multirow{-5}{*}{\raggedright\arraybackslash Other} & CHN &  & 1\\
\hline
\end{tabular}
\end{table}

\newpage

\begin{figure}

{\centering \includegraphics[width=1\linewidth,]{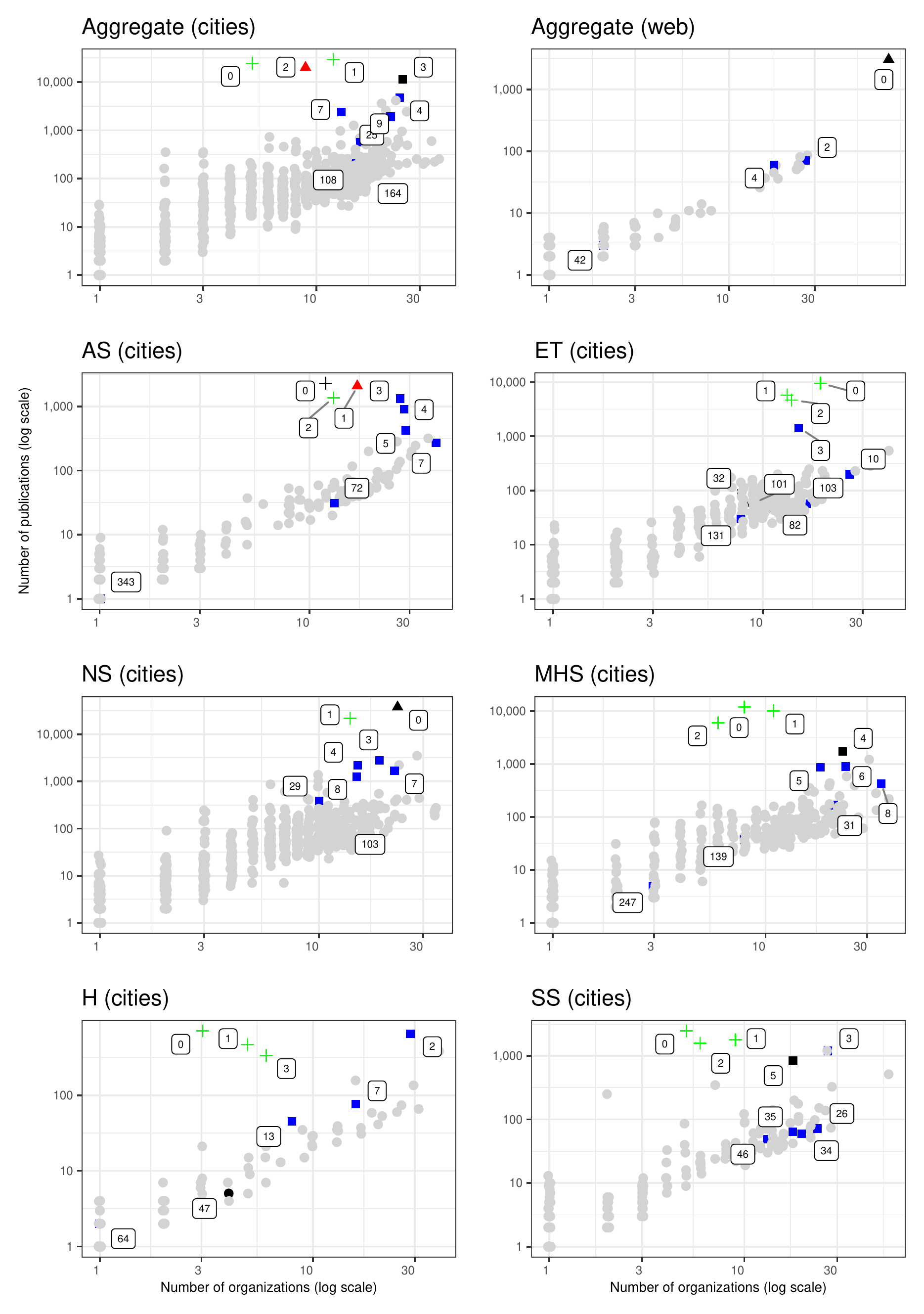} 

}

\caption{Organizations/publications of communities (label: name of community, black: includes iDiv center, red/triangle: includes Unibund and iDiv members, blue/square: includes only idiv member(s), green/plus: includes only Unibund members, gray/circle: others)}\label{fig:composition-of-clusters}
\end{figure}

Figure \ref{fig:composition-of-clusters} presents the communities identified from the giant component of iDiv web and HJL cities (\textbf{RQ5-8}). In the case of cities, it presents scientific fields separately. Note that in both cases, all Unibund and iDiv members are part of the giant component which is connected in itself. But, some institutions formed additional collaboration ties with one another, leading to denser areas in the giant component. Our first aim was to identify these denser areas and the underlying factors behind these higher densities and groupings. The labels printed beside some of the dots show the names of the communities. Each dot or shape shows one community including multiple institutions. The count of institutions in each community is indicated on the X axis which has a log scale. The Y axis shows the aggregate number of publications by those institutions in a community on a log scale. The gray circles show communities which did not contain an Unibund or iDiv member. If a community includes only Unibund members, it is indicated with a \emph{green plus}, and if it includes only iDiv members, a \emph{blue square} is used. When the iDiv center itself is in a given community, the color is set to \emph{black}. When a community includes a combination of iDiv and Unibund members, color is set to \emph{red} and shape is set to \emph{triangle}. Therefore, if high cohesiveness exists, we expect to see a black triangle.

In the case of iDiv publications (i.e., web), shown in the uppermost right-hand panel, except for three communities with iDiv members (i.e., blue squares), all other institutions (i.e., iDiv center itself, Unibund and iDiv members) are located in the most prolific community, i.e., 0, which indicates a high cohesion in collaborations. When the larger context of the HJL cities is considered, which still includes iDiv and Unibund members but they are placed in the regional context along with other institutions and their collaborators worldwide, then the identified denser areas of the giant component are further apart from each other. This is presented in the aggregate of cities in the uppermost left-hand panel where there are two green pluses with Unibund members, a triangle with an Unibund member and one iDiv member, versus multiple communities with iDiv members. We present the variation among scientific fields only in the case of the cities, since for iDiv web, only a small number of publications (7\%) were assigned to fields other than NS. The structure of collaborations is closest between cities and iDiv web, only in the case of NS which is the main field of publication for iDiv and to a lesser extent in AS. But, even in NS there are six separate communities with iDiv members and iDiv center is located in community 0 with two Unibund members. In all other fields, Unibund and iDiv members are located in separate communities, which indicates a lower cohesive collaboration structure in contrast to iDiv web sample. As we discuss next, it can also point to the disciplinary strength and scientific focus of these institutions which leads to the formation of a distinct group of regional, national or global collaborators that goes beyond the coalition agreements in the framework of Unibund or iDiv.

Table \ref{tab:composition-of-clusters-of-continents-ror-web-table} presents more detail on the communities (\textbf{RQ5-8}) that include Unibund and/or iDiv members (i.e., the communities indicated with labels on figure \ref{fig:composition-of-clusters}). It highlights the diversity of community members based on the percentage share of geographical regions and sectors. In geographical regions we differentiate between HJL cities, Berlin, Germany and Europe to differentiate local, national, continent or global based collaborations. In case of iDiv publications (i.e., Aggregate web), community 0 has 77 member institutions, includes all Unibund and five of iDiv members, plus the iDiv center itself. It also shows a high rate of geographical diversity with members coming from all regions. But, in the larger context (e.g., cities), iDiv is located in community 3 with only one other iDiv member institute. The geographical composition of this community is mainly focused on Germany, Europe and HJL cities with only 4\% of members from Oceania which is a highly different composition in contrast to community 0 in the iDiv web. Other iDiv and Unibund members are located in separate communities. While communities with Unibund members are less geographically diverse and mainly focused in Germany and Europe, communities with iDiv members have more international members. In terms of sectors, the majority of the community members are composed of education and facility. Government institutions are present only in communities with iDiv members. Health-care and non-profits are present in only a few of the communities. In the scientific fields, on the one hand, in most cases communities including Unibund members are composed of local and European institutions. Exceptions include four out of 17 communities i.e., community 1 in AS, community 0 in ET, community 1 in NS and community 3 in H where Unibund members have collaborated with institutions from other geographical regions. On the other hand, all of the communities with iDiv members are composed of international and global institutions. Communities where iDiv center is located (see rows with bold font and gray background) follow the latter rule, but to a lesser extent.

\begin{landscape}\begin{table}

\caption{\label{tab:composition-of-clusters-of-continents-ror-web-table}Composition of the communities including Unibund and/or iDiv members by region and sector in aggregate and by fields (N = community size, P = aggregate publications, iDiv = Y; includes iDiv center, I = number of iDiv members, U = number of Unibund members)}
\centering
\resizebox{\linewidth}{!}{
\begin{tabular}[t]{l|r|r|r|l|r|r|r|r|r|r|r|r|r|r|r|r|r|r|r|r|r|r|r}
\hline
\multicolumn{7}{c|}{ } & \multicolumn{9}{c|}{Region (\%)} & \multicolumn{8}{c}{Sector (\%)} \\
\cline{8-16} \cline{17-24}
\cellcolor{lightgray}{\textcolor{black}{\textbf{Data}}} & \cellcolor{lightgray}{\textcolor{black}{\textbf{cluster}}} & \cellcolor{lightgray}{\textcolor{black}{\textbf{N}}} & \cellcolor{lightgray}{\textcolor{black}{\textbf{P}}} & \cellcolor{lightgray}{\textcolor{black}{\textbf{iDiv}}} & \cellcolor{lightgray}{\textcolor{black}{\textbf{I}}} & \cellcolor{lightgray}{\textcolor{black}{\textbf{U}}} & \cellcolor{lightgray}{\textcolor{black}{\textbf{Africa}}} & \cellcolor{lightgray}{\textcolor{black}{\textbf{Americas}}} & \cellcolor{lightgray}{\textcolor{black}{\textbf{Asia}}} & \cellcolor{lightgray}{\textcolor{black}{\textbf{Berlin}}} & \cellcolor{lightgray}{\textcolor{black}{\textbf{DEU}}} & \cellcolor{lightgray}{\textcolor{black}{\textbf{Europe}}} & \cellcolor{lightgray}{\textcolor{black}{\textbf{HJL cities}}} & \cellcolor{lightgray}{\textcolor{black}{\textbf{Oceania}}} & \cellcolor{lightgray}{\textcolor{black}{\textbf{No region}}} & \cellcolor{lightgray}{\textcolor{black}{\textbf{Archive}}} & \cellcolor{lightgray}{\textcolor{black}{\textbf{Company}}} & \cellcolor{lightgray}{\textcolor{black}{\textbf{Education}}} & \cellcolor{lightgray}{\textcolor{black}{\textbf{Facility}}} & \cellcolor{lightgray}{\textcolor{black}{\textbf{Government}}} & \cellcolor{lightgray}{\textcolor{black}{\textbf{Healthcare}}} & \cellcolor{lightgray}{\textcolor{black}{\textbf{Nonprofit}}} & \cellcolor{lightgray}{\textcolor{black}{\textbf{Other}}}\\
\hline
 & 0 & 5 & 24,616 &  &  & 1 &  &  &  & 20 & 20 &  & 60 &  &  &  &  & 60 & 20 &  & 20 &  & \\

 & 1 & 12 & 29,608 &  &  & 1 &  &  &  &  & 25 & 17 & 50 & 8 &  &  &  & 42 & 42 &  & 17 &  & \\

 & 2 & 9 & 20,127 &  & 1 & 1 &  &  &  &  & 67 &  & 33 &  &  &  &  & 67 & 33 &  &  &  & \\

\cellcolor{lightgray}{\textcolor{black}{\textbf{}}} & \cellcolor{lightgray}{\textcolor{black}{\textbf{3}}} & \cellcolor{lightgray}{\textcolor{black}{\textbf{25}}} & \cellcolor{lightgray}{\textcolor{black}{\textbf{11,481}}} & \cellcolor{lightgray}{\textcolor{black}{\textbf{Y}}} & \cellcolor{lightgray}{\textcolor{black}{\textbf{1}}} & \cellcolor{lightgray}{\textcolor{black}{\textbf{}}} & \cellcolor{lightgray}{\textcolor{black}{\textbf{}}} & \cellcolor{lightgray}{\textcolor{black}{\textbf{}}} & \cellcolor{lightgray}{\textcolor{black}{\textbf{}}} & \cellcolor{lightgray}{\textcolor{black}{\textbf{}}} & \cellcolor{lightgray}{\textcolor{black}{\textbf{48}}} & \cellcolor{lightgray}{\textcolor{black}{\textbf{36}}} & \cellcolor{lightgray}{\textcolor{black}{\textbf{12}}} & \cellcolor{lightgray}{\textcolor{black}{\textbf{4}}} & \cellcolor{lightgray}{\textcolor{black}{\textbf{}}} & \cellcolor{lightgray}{\textcolor{black}{\textbf{}}} & \cellcolor{lightgray}{\textcolor{black}{\textbf{}}} & \cellcolor{lightgray}{\textcolor{black}{\textbf{72}}} & \cellcolor{lightgray}{\textcolor{black}{\textbf{20}}} & \cellcolor{lightgray}{\textcolor{black}{\textbf{4}}} & \cellcolor{lightgray}{\textcolor{black}{\textbf{}}} & \cellcolor{lightgray}{\textcolor{black}{\textbf{4}}} & \cellcolor{lightgray}{\textcolor{black}{\textbf{}}}\\

 & 4 & 24 & 4,741 &  & 1 &  &  & 42 &  &  & 8 & 46 & 4 &  &  &  &  & 79 & 17 & 4 &  &  & \\

 & 7 & 13 & 2,398 &  & 1 &  &  & 31 & 15 &  &  & 46 & 8 &  &  &  &  & 62 & 31 & 8 &  &  & \\

 & 9 & 22 & 1,912 &  & 1 &  & 5 & 14 &  &  & 27 & 50 & 5 &  &  &  &  & 27 & 45 & 14 &  & 9 & 5\\

 & 25 & 16 & 569 &  & 1 &  & 6 & 6 & 12 &  & 44 & 31 &  &  &  &  & 6 & 56 & 19 & 6 & 12 &  & \\

 & 108 & 15 & 208 &  & 1 &  & 7 &  & 27 &  & 27 & 40 &  &  &  &  &  & 47 & 33 & 7 &  & 7 & 7\\

\multirow{-10}{*}{\raggedright\arraybackslash Aggregate cities} & 164 & 17 & 109 &  & 1 &  &  & 6 & 6 &  & 12 & 76 &  &  &  & 18 & 6 & 29 & 35 & 6 &  &  & 6\\
\addlinespace
\cellcolor{lightgray}{\textcolor{black}{\textbf{}}} & \cellcolor{lightgray}{\textcolor{black}{\textbf{0}}} & \cellcolor{lightgray}{\textcolor{black}{\textbf{77}}} & \cellcolor{lightgray}{\textcolor{black}{\textbf{3,061}}} & \cellcolor{lightgray}{\textcolor{black}{\textbf{Y}}} & \cellcolor{lightgray}{\textcolor{black}{\textbf{5}}} & \cellcolor{lightgray}{\textcolor{black}{\textbf{3}}} & \cellcolor{lightgray}{\textcolor{black}{\textbf{3}}} & \cellcolor{lightgray}{\textcolor{black}{\textbf{5}}} & \cellcolor{lightgray}{\textcolor{black}{\textbf{9}}} & \cellcolor{lightgray}{\textcolor{black}{\textbf{4}}} & \cellcolor{lightgray}{\textcolor{black}{\textbf{38}}} & \cellcolor{lightgray}{\textcolor{black}{\textbf{29}}} & \cellcolor{lightgray}{\textcolor{black}{\textbf{10}}} & \cellcolor{lightgray}{\textcolor{black}{\textbf{3}}} & \cellcolor{lightgray}{\textcolor{black}{\textbf{}}} & \cellcolor{lightgray}{\textcolor{black}{\textbf{1}}} & \cellcolor{lightgray}{\textcolor{black}{\textbf{}}} & \cellcolor{lightgray}{\textcolor{black}{\textbf{70}}} & \cellcolor{lightgray}{\textcolor{black}{\textbf{22}}} & \cellcolor{lightgray}{\textcolor{black}{\textbf{3}}} & \cellcolor{lightgray}{\textcolor{black}{\textbf{}}} & \cellcolor{lightgray}{\textcolor{black}{\textbf{3}}} & \cellcolor{lightgray}{\textcolor{black}{\textbf{1}}}\\

 & 2 & 27 & 71 &  & 1 &  & 4 &  & 15 &  & 22 & 59 &  &  &  & 7 &  & 63 & 7 & 22 &  &  & \\

 & 4 & 18 & 60 &  & 1 &  & 33 & 17 &  &  & 11 & 33 & 6 &  &  & 6 & 6 & 50 & 11 & 6 &  & 11 & 11\\

\multirow{-4}{*}{\raggedright\arraybackslash Aggregate web} & 42 & 2 & 3 &  & 1 &  &  &  &  &  & 100 &  &  &  &  &  &  &  & 100 &  &  &  & \\
\addlinespace
\cellcolor{lightgray}{\textcolor{black}{\textbf{}}} & \cellcolor{lightgray}{\textcolor{black}{\textbf{0}}} & \cellcolor{lightgray}{\textcolor{black}{\textbf{12}}} & \cellcolor{lightgray}{\textcolor{black}{\textbf{2,298}}} & \cellcolor{lightgray}{\textcolor{black}{\textbf{Y}}} & \cellcolor{lightgray}{\textcolor{black}{\textbf{}}} & \cellcolor{lightgray}{\textcolor{black}{\textbf{1}}} & \cellcolor{lightgray}{\textcolor{black}{\textbf{}}} & \cellcolor{lightgray}{\textcolor{black}{\textbf{8}}} & \cellcolor{lightgray}{\textcolor{black}{\textbf{}}} & \cellcolor{lightgray}{\textcolor{black}{\textbf{17}}} & \cellcolor{lightgray}{\textcolor{black}{\textbf{33}}} & \cellcolor{lightgray}{\textcolor{black}{\textbf{25}}} & \cellcolor{lightgray}{\textcolor{black}{\textbf{17}}} & \cellcolor{lightgray}{\textcolor{black}{\textbf{}}} & \cellcolor{lightgray}{\textcolor{black}{\textbf{}}} & \cellcolor{lightgray}{\textcolor{black}{\textbf{}}} & \cellcolor{lightgray}{\textcolor{black}{\textbf{}}} & \cellcolor{lightgray}{\textcolor{black}{\textbf{67}}} & \cellcolor{lightgray}{\textcolor{black}{\textbf{8}}} & \cellcolor{lightgray}{\textcolor{black}{\textbf{8}}} & \cellcolor{lightgray}{\textcolor{black}{\textbf{8}}} & \cellcolor{lightgray}{\textcolor{black}{\textbf{}}} & \cellcolor{lightgray}{\textcolor{black}{\textbf{8}}}\\

 & 1 & 17 & 2,086 &  & 2 & 1 & 18 &  &  &  & 71 &  & 12 &  &  &  &  & 59 & 35 & 6 &  &  & \\

 & 2 & 13 & 1,363 &  &  & 1 &  & 8 &  &  & 54 &  & 38 &  &  &  &  & 62 & 15 & 15 & 8 &  & \\

 & 3 & 27 & 1,325 &  & 1 &  &  & 11 & 4 & 4 & 48 & 19 & 11 & 4 &  &  &  & 67 & 30 &  &  & 4 & \\

 & 4 & 28 & 903 &  & 1 &  &  & 39 & 4 & 7 & 7 & 39 & 4 &  &  &  &  & 71 & 18 & 7 &  & 4 & \\

 & 5 & 29 & 426 &  & 1 &  & 3 & 31 & 17 &  & 17 & 28 & 3 &  &  &  & 3 & 62 & 28 & 3 &  &  & 3\\

 & 7 & 40 & 271 &  & 1 &  & 25 & 10 & 12 &  & 5 & 40 & 8 &  &  &  & 2 & 52 & 28 & 2 & 2 & 10 & 2\\

 & 72 & 13 & 31 &  & 1 &  &  & 38 & 8 &  & 15 & 38 &  &  &  & 8 &  & 38 & 31 & 15 &  & 8 & \\

\multirow{-9}{*}{\raggedright\arraybackslash AS cities} & 343 & 1 & 1 &  & 1 &  &  &  &  &  & 100 &  &  &  &  &  &  &  & 100 &  &  &  & \\
\addlinespace
 & 0 & 19 & 9,540 &  &  & 1 &  & 5 & 5 &  & 42 & 11 & 32 & 5 &  &  & 11 & 42 & 37 &  & 11 &  & \\

 & 1 & 13 & 5,773 &  &  & 1 &  &  &  & 8 & 54 & 23 & 15 &  &  &  &  & 54 & 38 &  &  & 8 & \\

 & 2 & 14 & 4,664 &  &  & 1 &  &  &  &  & 50 & 29 & 21 &  &  &  &  & 64 & 21 & 7 & 7 &  & \\

 & 3 & 15 & 1,420 &  & 1 &  &  &  &  & 7 & 47 & 20 & 20 & 7 &  &  &  & 53 & 33 & 7 &  & 7 & \\

 & 10 & 26 & 200 &  & 1 &  & 4 & 27 & 4 &  & 4 & 58 & 4 &  &  &  & 4 & 69 & 15 & 4 & 4 & 4 & \\

 & 32 & 8 & 89 &  & 1 &  &  &  & 25 &  &  & 62 & 12 &  &  &  &  & 62 & 38 &  &  &  & \\

 & 82 & 11 & 46 &  & 1 &  & 9 & 18 & 36 &  & 9 & 18 & 9 &  &  &  & 9 & 73 & 18 &  &  &  & \\

\cellcolor{lightgray}{\textcolor{black}{\textbf{}}} & \cellcolor{lightgray}{\textcolor{black}{\textbf{101}}} & \cellcolor{lightgray}{\textcolor{black}{\textbf{9}}} & \cellcolor{lightgray}{\textcolor{black}{\textbf{58}}} & \cellcolor{lightgray}{\textcolor{black}{\textbf{Y}}} & \cellcolor{lightgray}{\textcolor{black}{\textbf{}}} & \cellcolor{lightgray}{\textcolor{black}{\textbf{}}} & \cellcolor{lightgray}{\textcolor{black}{\textbf{}}} & \cellcolor{lightgray}{\textcolor{black}{\textbf{}}} & \cellcolor{lightgray}{\textcolor{black}{\textbf{44}}} & \cellcolor{lightgray}{\textcolor{black}{\textbf{}}} & \cellcolor{lightgray}{\textcolor{black}{\textbf{11}}} & \cellcolor{lightgray}{\textcolor{black}{\textbf{33}}} & \cellcolor{lightgray}{\textcolor{black}{\textbf{11}}} & \cellcolor{lightgray}{\textcolor{black}{\textbf{}}} & \cellcolor{lightgray}{\textcolor{black}{\textbf{}}} & \cellcolor{lightgray}{\textcolor{black}{\textbf{}}} & \cellcolor{lightgray}{\textcolor{black}{\textbf{}}} & \cellcolor{lightgray}{\textcolor{black}{\textbf{67}}} & \cellcolor{lightgray}{\textcolor{black}{\textbf{22}}} & \cellcolor{lightgray}{\textcolor{black}{\textbf{11}}} & \cellcolor{lightgray}{\textcolor{black}{\textbf{}}} & \cellcolor{lightgray}{\textcolor{black}{\textbf{}}} & \cellcolor{lightgray}{\textcolor{black}{\textbf{}}}\\

 & 103 & 16 & 57 &  & 1 &  &  & 38 & 25 &  &  & 25 & 6 & 6 &  &  &  & 56 & 38 & 6 &  &  & \\

\multirow{-10}{*}{\raggedright\arraybackslash ET cities} & 131 & 8 & 30 &  & 1 &  &  & 50 & 25 &  & 25 &  &  &  &  &  & 12 & 50 & 12 & 12 & 12 &  & \\
\addlinespace
\cellcolor{lightgray}{\textcolor{black}{\textbf{}}} & \cellcolor{lightgray}{\textcolor{black}{\textbf{0}}} & \cellcolor{lightgray}{\textcolor{black}{\textbf{23}}} & \cellcolor{lightgray}{\textcolor{black}{\textbf{37,765}}} & \cellcolor{lightgray}{\textcolor{black}{\textbf{Y}}} & \cellcolor{lightgray}{\textcolor{black}{\textbf{1}}} & \cellcolor{lightgray}{\textcolor{black}{\textbf{2}}} & \cellcolor{lightgray}{\textcolor{black}{\textbf{}}} & \cellcolor{lightgray}{\textcolor{black}{\textbf{}}} & \cellcolor{lightgray}{\textcolor{black}{\textbf{}}} & \cellcolor{lightgray}{\textcolor{black}{\textbf{9}}} & \cellcolor{lightgray}{\textcolor{black}{\textbf{57}}} & \cellcolor{lightgray}{\textcolor{black}{\textbf{9}}} & \cellcolor{lightgray}{\textcolor{black}{\textbf{26}}} & \cellcolor{lightgray}{\textcolor{black}{\textbf{}}} & \cellcolor{lightgray}{\textcolor{black}{\textbf{}}} & \cellcolor{lightgray}{\textcolor{black}{\textbf{}}} & \cellcolor{lightgray}{\textcolor{black}{\textbf{}}} & \cellcolor{lightgray}{\textcolor{black}{\textbf{78}}} & \cellcolor{lightgray}{\textcolor{black}{\textbf{22}}} & \cellcolor{lightgray}{\textcolor{black}{\textbf{}}} & \cellcolor{lightgray}{\textcolor{black}{\textbf{}}} & \cellcolor{lightgray}{\textcolor{black}{\textbf{}}} & \cellcolor{lightgray}{\textcolor{black}{\textbf{}}}\\

 & 1 & 14 & 21,880 &  &  & 1 &  & 7 & 7 &  & 29 & 14 & 36 & 7 &  &  & 14 & 43 & 36 &  & 7 &  & \\

 & 3 & 19 & 2,785 &  & 1 &  &  & 37 &  &  & 11 & 47 & 5 &  &  &  &  & 79 & 21 &  &  &  & \\

 & 4 & 15 & 2,191 &  & 1 &  & 7 & 13 & 20 &  & 13 & 33 & 7 & 7 &  &  &  & 60 & 40 &  &  &  & \\

 & 7 & 22 & 1,694 &  & 1 &  & 5 & 23 &  &  & 14 & 45 & 5 & 9 &  &  & 5 & 32 & 36 & 18 &  & 5 & 5\\

 & 8 & 15 & 1,253 &  & 2 &  & 13 & 33 &  &  & 33 & 13 & 7 &  &  &  &  & 40 & 53 &  &  & 7 & \\

 & 29 & 10 & 384 &  & 1 &  &  &  & 10 &  & 70 & 20 &  &  &  & 10 & 20 & 40 & 10 &  & 10 & 10 & \\

\multirow{-8}{*}{\raggedright\arraybackslash NS cities} & 103 & 13 & 105 &  & 1 &  &  & 8 & 38 &  & 15 & 38 &  &  &  &  &  & 62 & 23 & 8 & 8 &  & \\
\addlinespace
 & 0 & 8 & 11,901 &  &  & 1 &  &  &  &  & 62 & 12 & 25 &  &  &  &  & 75 &  &  & 25 &  & \\

 & 1 & 11 & 10,104 &  &  & 1 &  &  &  &  & 55 &  & 45 &  &  &  &  & 55 & 18 &  & 27 &  & \\

 & 2 & 6 & 6,038 &  &  & 1 &  &  &  &  & 67 &  & 33 &  &  &  &  & 67 &  &  & 33 &  & \\

\cellcolor{lightgray}{\textcolor{black}{\textbf{}}} & \cellcolor{lightgray}{\textcolor{black}{\textbf{4}}} & \cellcolor{lightgray}{\textcolor{black}{\textbf{23}}} & \cellcolor{lightgray}{\textcolor{black}{\textbf{1,724}}} & \cellcolor{lightgray}{\textcolor{black}{\textbf{Y}}} & \cellcolor{lightgray}{\textcolor{black}{\textbf{1}}} & \cellcolor{lightgray}{\textcolor{black}{\textbf{}}} & \cellcolor{lightgray}{\textcolor{black}{\textbf{}}} & \cellcolor{lightgray}{\textcolor{black}{\textbf{4}}} & \cellcolor{lightgray}{\textcolor{black}{\textbf{}}} & \cellcolor{lightgray}{\textcolor{black}{\textbf{}}} & \cellcolor{lightgray}{\textcolor{black}{\textbf{39}}} & \cellcolor{lightgray}{\textcolor{black}{\textbf{43}}} & \cellcolor{lightgray}{\textcolor{black}{\textbf{13}}} & \cellcolor{lightgray}{\textcolor{black}{\textbf{}}} & \cellcolor{lightgray}{\textcolor{black}{\textbf{}}} & \cellcolor{lightgray}{\textcolor{black}{\textbf{}}} & \cellcolor{lightgray}{\textcolor{black}{\textbf{4}}} & \cellcolor{lightgray}{\textcolor{black}{\textbf{61}}} & \cellcolor{lightgray}{\textcolor{black}{\textbf{22}}} & \cellcolor{lightgray}{\textcolor{black}{\textbf{}}} & \cellcolor{lightgray}{\textcolor{black}{\textbf{4}}} & \cellcolor{lightgray}{\textcolor{black}{\textbf{9}}} & \cellcolor{lightgray}{\textcolor{black}{\textbf{}}}\\

 & 5 & 18 & 862 &  & 1 &  & 6 & 22 & 28 &  &  & 33 & 6 & 6 &  &  &  & 67 & 17 & 11 &  &  & 6\\

 & 6 & 24 & 894 &  & 1 &  &  & 58 & 8 &  & 4 & 25 & 4 &  &  &  &  & 75 & 21 &  &  &  & 4\\

 & 8 & 35 & 429 &  & 2 &  & 6 & 34 & 20 & 3 & 11 & 23 & 3 &  &  & 3 &  & 57 & 34 &  & 6 &  & \\

 & 31 & 21 & 167 &  & 1 &  &  & 57 & 14 &  & 10 & 14 & 5 &  &  &  &  & 62 & 24 & 14 &  &  & \\

 & 139 & 8 & 37 &  & 1 &  &  & 12 & 12 &  & 12 & 62 &  &  &  & 12 &  & 50 & 25 & 12 &  &  & \\

\multirow{-10}{*}{\raggedright\arraybackslash MHS cities} & 247 & 3 & 5 &  & 1 &  &  &  & 67 &  & 33 &  &  &  &  & 33 &  & 67 &  &  &  &  & \\
\addlinespace
 & 0 & 3 & 707 &  &  & 1 &  &  &  &  & 33 & 33 & 33 &  &  &  &  & 67 & 33 &  &  &  & \\

 & 1 & 5 & 468 &  &  & 1 &  &  & 20 &  & 60 &  & 20 &  &  &  &  & 80 &  & 20 &  &  & \\

 & 2 & 29 & 657 &  & 1 &  & 3 & 34 &  &  & 7 & 41 & 3 & 7 & 3 &  & 3 & 90 & 7 &  &  &  & \\

 & 3 & 6 & 337 &  &  & 1 &  &  & 17 &  & 50 & 17 & 17 &  &  &  &  & 100 &  &  &  &  & \\

 & 7 & 16 & 77 &  & 1 &  &  & 6 &  & 6 & 44 & 25 & 12 &  & 6 & 19 &  & 56 & 19 &  &  & 6 & \\

 & 13 & 8 & 45 &  & 1 &  & 12 &  & 12 &  &  & 50 & 25 &  &  &  &  & 62 & 38 &  &  &  & \\

\cellcolor{lightgray}{\textcolor{black}{\textbf{}}} & \cellcolor{lightgray}{\textcolor{black}{\textbf{47}}} & \cellcolor{lightgray}{\textcolor{black}{\textbf{4}}} & \cellcolor{lightgray}{\textcolor{black}{\textbf{5}}} & \cellcolor{lightgray}{\textcolor{black}{\textbf{Y}}} & \cellcolor{lightgray}{\textcolor{black}{\textbf{}}} & \cellcolor{lightgray}{\textcolor{black}{\textbf{}}} & \cellcolor{lightgray}{\textcolor{black}{\textbf{}}} & \cellcolor{lightgray}{\textcolor{black}{\textbf{}}} & \cellcolor{lightgray}{\textcolor{black}{\textbf{}}} & \cellcolor{lightgray}{\textcolor{black}{\textbf{}}} & \cellcolor{lightgray}{\textcolor{black}{\textbf{}}} & \cellcolor{lightgray}{\textcolor{black}{\textbf{75}}} & \cellcolor{lightgray}{\textcolor{black}{\textbf{25}}} & \cellcolor{lightgray}{\textcolor{black}{\textbf{}}} & \cellcolor{lightgray}{\textcolor{black}{\textbf{}}} & \cellcolor{lightgray}{\textcolor{black}{\textbf{}}} & \cellcolor{lightgray}{\textcolor{black}{\textbf{}}} & \cellcolor{lightgray}{\textcolor{black}{\textbf{25}}} & \cellcolor{lightgray}{\textcolor{black}{\textbf{50}}} & \cellcolor{lightgray}{\textcolor{black}{\textbf{25}}} & \cellcolor{lightgray}{\textcolor{black}{\textbf{}}} & \cellcolor{lightgray}{\textcolor{black}{\textbf{}}} & \cellcolor{lightgray}{\textcolor{black}{\textbf{}}}\\

\multirow{-8}{*}{\raggedright\arraybackslash H cities} & 64 & 1 & 2 &  & 1 &  &  &  &  &  &  &  & 100 &  &  &  &  &  & 100 &  &  &  & \\
\addlinespace
 & 0 & 5 & 2,465 &  &  & 1 &  &  &  &  & 40 & 20 & 40 &  &  &  &  & 60 &  &  & 40 &  & \\

 & 1 & 9 & 1,787 &  &  & 1 &  &  &  &  & 78 & 11 & 11 &  &  &  &  & 100 &  &  &  &  & \\

 & 2 & 6 & 1,580 &  &  & 1 &  &  &  &  & 67 &  & 33 &  &  & 17 &  & 83 &  &  &  &  & \\

 & 3 & 27 & 1,197 &  & 1 &  & 4 & 37 & 4 &  & 7 & 41 & 4 &  & 4 & 7 &  & 85 & 7 &  &  &  & \\

\cellcolor{lightgray}{\textcolor{black}{\textbf{}}} & \cellcolor{lightgray}{\textcolor{black}{\textbf{5}}} & \cellcolor{lightgray}{\textcolor{black}{\textbf{18}}} & \cellcolor{lightgray}{\textcolor{black}{\textbf{843}}} & \cellcolor{lightgray}{\textcolor{black}{\textbf{Y}}} & \cellcolor{lightgray}{\textcolor{black}{\textbf{1}}} & \cellcolor{lightgray}{\textcolor{black}{\textbf{}}} & \cellcolor{lightgray}{\textcolor{black}{\textbf{}}} & \cellcolor{lightgray}{\textcolor{black}{\textbf{6}}} & \cellcolor{lightgray}{\textcolor{black}{\textbf{}}} & \cellcolor{lightgray}{\textcolor{black}{\textbf{11}}} & \cellcolor{lightgray}{\textcolor{black}{\textbf{44}}} & \cellcolor{lightgray}{\textcolor{black}{\textbf{17}}} & \cellcolor{lightgray}{\textcolor{black}{\textbf{17}}} & \cellcolor{lightgray}{\textcolor{black}{\textbf{6}}} & \cellcolor{lightgray}{\textcolor{black}{\textbf{}}} & \cellcolor{lightgray}{\textcolor{black}{\textbf{}}} & \cellcolor{lightgray}{\textcolor{black}{\textbf{}}} & \cellcolor{lightgray}{\textcolor{black}{\textbf{61}}} & \cellcolor{lightgray}{\textcolor{black}{\textbf{22}}} & \cellcolor{lightgray}{\textcolor{black}{\textbf{}}} & \cellcolor{lightgray}{\textcolor{black}{\textbf{}}} & \cellcolor{lightgray}{\textcolor{black}{\textbf{11}}} & \cellcolor{lightgray}{\textcolor{black}{\textbf{6}}}\\

 & 26 & 24 & 72 &  & 1 &  &  & 58 & 4 &  &  & 33 & 4 &  &  &  &  & 62 & 21 & 4 & 12 &  & \\

 & 34 & 20 & 60 &  & 1 &  &  & 10 & 30 &  & 5 & 45 & 5 & 5 &  &  &  & 80 & 10 & 5 & 5 &  & \\

 & 35 & 18 & 64 &  & 1 &  & 11 & 22 & 11 &  & 6 & 22 & 6 & 22 &  & 6 &  & 39 & 22 & 11 & 17 & 6 & \\

\multirow{-9}{*}{\raggedright\arraybackslash SS cities} & 46 & 13 & 49 &  & 1 &  & 8 & 54 & 8 &  & 8 & 23 &  &  &  & 8 &  & 85 &  & 8 &  &  & \\
\hline
\end{tabular}}
\end{table}
\end{landscape}

\hypertarget{conclusions}{%
\section{Discussion and Conclusions}\label{conclusions}}

This paper quantitatively explored the structure of institutional scientific collaborations in a geographical region in the central Germany comprised of three cities, Halle (Saale), Jena, and Leipzig. The three universities of these cities joined forces to form a strategic coalition, i.e., Unibund, that has lasted 25 years. In addition, this is an interesting region since it does not comply with the idea of a centralized and concentrated metropolitan area. From 2012, this coalition, in collaboration with Helmholtz Centre for Environmental Research (UFZ) and seven other non-university institutions has established a new organizational entity, the German Center for Integrative Biodiversity Research (iDiv). Our main intent was to explore how Unibund members have collaborated with each other and to what extent iDiv as the new organizational form has integrated its diverse group of members into the collaboration network. In addition, we compared the structure of scientific collaborations based on the self-represented scientific output of iDiv on their website with the larger context of scientific output of the HJL region.

First, in methodological terms, we presented the effects of disambiguation of organization names on results. We showed that disambiguation can not be overlooked if we intend to construct networks and investigate collaboration trajectories. Any attempt at doing so without proper disambiguation would be reductionist and erroneous (see \protect\hyperlink{ref-akbaritabarBerlinQuantitativeView2020}{Akbaritabar} (\protect\hyperlink{ref-akbaritabarBerlinQuantitativeView2020}{2020}) for further discussion).

Moreover, the German science system in general (e.g., \protect\hyperlink{ref-stahlschmidtPerformanceStructuresGerman2019}{Stahlschmidt, Stephen, \& Hinze} (\protect\hyperlink{ref-stahlschmidtPerformanceStructuresGerman2019}{2019}); \protect\hyperlink{ref-stephenPerformanceStructuresGerman2020}{Stephen, Stahlschmidt, \& Hinze} (\protect\hyperlink{ref-stephenPerformanceStructuresGerman2020}{2020})) and HJL region specifically, present a stable trend of fractional counts of publications versus raw counts indicating a mature trend of institutional co-authorship and team science. But, in the case of the disambiguated publications studied here, we observe that most co-authorship occurs within Germany and only a few fields (e.g., NS and AS and to a lesser extent ET) show higher rates of multiple country co-authorship which is similar to \protect\hyperlink{ref-akbaritabarBerlinQuantitativeView2020}{Akbaritabar} (\protect\hyperlink{ref-akbaritabarBerlinQuantitativeView2020}{2020})'s findings of the Berlin metropolitan region. Note that this is in line with the focus of natural versus social sciences, where natural sciences share resources over large, global teams to manage costs and work on globally relevant research questions, whereas more social sciences and humanities work focuses on regional/local questions and is conducted by smaller teams. However, exceptional cases might exist to these disciplinary traditions.

Research activity in many countries is still dominated by the education sector (or in rare cases such as Germany, facility and education dominate the picture) and other sectors have a long way to go to catch up. There is still much to be done for university-industry relations and companies (as representative of business sector) have only marginal positions in a few small and less prolific clusters and their collaboration is limited to specific fields (e.g., AS, ET and NS).

Regarding the main aims of our paper, we found that Unibund has mainly persisted as a top-down policy inspiring further scientific collaboration. But in practice, while its members collaborate in the framework of the coalition, they maintain a more cohesive and distinctive group of collaborators distributed nationally or in Europe. iDiv as an interdisciplinary network has been successfully established. Based on the self-represented scientific output and in only six years covered in our study, it has attracted and integrated a diverse group of institutions as members. Furthermore, it has established international and global collaboration ties. But, once we compared structure of collaborations \emph{within} and \emph{between} Unibund and iDiv members and scientific output of the larger context of HJL region, we observed that the highly cohesive structure based on iDiv's publications was more the exception than the rule. Even though all three of Unibund and five of iDiv members were cohesively collaborating with each other in the community 0 based on publications extracted from iDiv web, based on the HJL cities' publications, only one iDiv member is located in the community 3 with iDiv center and other members are located in separate cohesive communities with a more (in case of iDiv members) or less (in case of Unibund members) international composition. iDiv members' collaboration with each other (i.e., the network constructed based on iDiv web data) is only a small part of their bigger collaboration trajectory. A larger share of their collaboration is still composed of working with a global group of other collaborators.

To conclude, although an \emph{Integrated European Research Area} (\protect\hyperlink{ref-hoekmanResearchCollaborationDistance2010}{Hoekman, Frenken, \& Tijssen, 2010}) has formed, nevertheless, geographical proximity (\protect\hyperlink{ref-ballandComplexEconomicActivities2020}{Balland et al., 2020}; \protect\hyperlink{ref-rammerKnowledgeProximityFirm2020}{Rammer, Kinne, \& Blind, 2020}) is not enough to ensure regional cooperation (\protect\hyperlink{ref-abbasiharoftehStillShadowWall2020}{Abbasiharofteh \& Broekel, 2020}) similar to top-down policies (\protect\hyperlink{ref-akbaritabarBerlinQuantitativeView2020}{Akbaritabar, 2020}). Furthermore, even use of \emph{bureaucracy} and establishment of new organizational entities (\protect\hyperlink{ref-shrumStructuresScientificCollaboration2007}{Shrum, Genuth, Carlson, Chompalov, \& Bijker, 2007} p 192, 200-201) do not ensure a cohesive collaboration structure. Unibund and iDiv have been relatively successful (e.g., in terms of number of publications) mainly in the natural sciences to shape the structure of scientific collaborations in the HJL region. But, in other scientific fields, where these institutions are actively publishing, a less cohesive collaboration structure is formed and there is room for these strategic coalitions to explore unrealized potentials.

The main limitation of our paper is the coverage of organization name disambiguation and being limited to only Scopus indexed content for scientific output. This does not include other modes of scientific collaboration that institutions might carry out such as sharing infrastructure and resources, to name a few. Our analysis is limited to the level of scientific organization and more detailed investigation in individual scientists level, which requires fine-grained author name disambiguation techniques, would reveal more reliable insight into the structure of scientific collaborations.

We modeled the networks as a bipartite one which is an improvement to the one-mode projecttion of it. But, the number of available algorithms to identify communities in this type of networks is still limited.Nevertheless, there are recent developments in the area of bipartite community detection and in our future work, we would like to evaluate robustness of our community detection results, using new algorithms such as \emph{BiMLPA} (\protect\hyperlink{ref-taguchiBiMLPACommunityDetection2020}{Taguchi, Murata, \& Liu, 2020}) with the standard implementaion of it in CDlib library (\protect\hyperlink{ref-rossettiCDLIBPythonLibrary2019}{Rossetti, Milli, \& Cazabet, 2019}) that allows for comparative analysis of community detection results. This will ensure that the structure we are observing is robust to multiple community detection methods and is an artefact of the scientific collaborations in the region and not a byproduct of the specific algorithm we used in our analysis.

\hypertarget{acknowledgements}{%
\section{Acknowledgements}\label{acknowledgements}}

We would like to thank Stephan Stahlschmidt, Dimity Stephen and Melike Janßen for comments and suggestions on earlier versions of this paper.

\hypertarget{funding-information}{%
\section{Funding Information}\label{funding-information}}

This research was done in DEKiF project supported by Federal Ministry for Education and Research (BMBF), Germany, with grant number: M527600. Data is obtained from Kompetenzzentrum Bibliometrie (Competence Center for Bibliometrics), Germany, which is funded by BMBF with grant number 01PQ17001.

\hypertarget{data-availability}{%
\section{Data Availability}\label{data-availability}}

Data cannot be made publicly available due to the licensing and contract terms of the original data.

\hypertarget{references}{%
\section*{References}\label{references}}
\addcontentsline{toc}{section}{References}

\hypertarget{refs}{}
\begin{CSLReferences}{1}{0}
\leavevmode\hypertarget{ref-abbasiharoftehStillShadowWall2020}{}%
Abbasiharofteh, M., \& Broekel, T. (2020). Still in the shadow of the wall? {The} case of the {Berlin} biotechnology cluster: \emph{Environment and Planning A: Economy and Space}. \url{https://doi.org/10.1177/0308518X20933904}

\leavevmode\hypertarget{ref-akbaritabarBerlinQuantitativeView2020}{}%
Akbaritabar, A. (2020). Berlin: {A Quantitative View} of the {Structure} of {Institutional Scientific Collaborations}. \emph{arXiv:2008.08355 {[}cs{]}}. Retrieved from \url{http://arxiv.org/abs/2008.08355}

\leavevmode\hypertarget{ref-ballandComplexEconomicActivities2020}{}%
Balland, P.-A., Jara-Figueroa, C., Petralia, S. G., Steijn, M. P. A., Rigby, D. L., \& Hidalgo, C. A. (2020). Complex economic activities concentrate in large cities. \emph{Nature Human Behaviour}, \emph{4}(3), 248--254. \url{https://doi.org/10.1038/s41562-019-0803-3}

\leavevmode\hypertarget{ref-biancaniSocialNetworksResearch2013}{}%
Biancani, S., \& McFarland, D. A. (2013). Social networks research in higher education. In \emph{Higher education: {Handbook} of theory and research} (pp. 151--215). {Springer}.

\leavevmode\hypertarget{ref-boshoffNeocolonialismResearchCollaboration2009}{}%
Boshoff, N. (2009). Neo-colonialism and research collaboration in {Central Africa}. \emph{Scientometrics}, \emph{81}(2), 413. \url{https://doi.org/10.1007/s11192-008-2211-8}

\leavevmode\hypertarget{ref-breigerDualityPersonsGroups1974}{}%
Breiger, R. L. (1974). The {Duality} of {Persons} and {Groups}. \emph{Social Forces}, \emph{53}(2), 181--190. \url{https://doi.org/10.1093/sf/53.2.181}

\leavevmode\hypertarget{ref-cravenEvolutionInterdisciplinarityBiodiversity2019}{}%
Craven, D., Winter, M., Hotzel, K., Gaikwad, J., Eisenhauer, N., Hohmuth, M., \ldots{} Wirth, C. (2019). Evolution of interdisciplinarity in biodiversity science. \emph{Ecology and Evolution}, \emph{9}(12), 6744--6755. \url{https://doi.org/10.1002/ece3.5244}

\leavevmode\hypertarget{ref-habelMoreEqualFooting2014}{}%
Habel, J. C., Eggermont, H., Günter, S., Mulwa, R. K., Rieckmann, M., Koh, L. P., \ldots{} Lens, L. (2014). Towards more equal footing in north{}south biodiversity research: {European} and sub-{Saharan} viewpoints. \emph{Biodiversity and Conservation}, \emph{23}(12), 3143--3148. \url{https://doi.org/10.1007/s10531-014-0761-z}

\leavevmode\hypertarget{ref-hoekmanResearchCollaborationDistance2010}{}%
Hoekman, J., Frenken, K., \& Tijssen, R. J. W. (2010). Research collaboration at a distance: {Changing} spatial patterns of scientific collaboration within {Europe}. \emph{Research Policy}, \emph{39}(5), 662--673. \url{https://doi.org/10.1016/j.respol.2010.01.012}

\leavevmode\hypertarget{ref-katzWhatResearchCollaboration1997}{}%
Katz, J. S., \& Martin, B. R. (1997). What is research collaboration? \emph{Research Policy}, \emph{26}(1), 1--18. \url{https://doi.org/10.1016/S0048-7333(96)00917-1}

\leavevmode\hypertarget{ref-newmanScientificCollaborationNetworks2001a}{}%
Newman, M. E. J. (2001a). Scientific collaboration networks. {I}. {Network} construction and fundamental results. \emph{Physical Review E}, \emph{64}(1), 016131. \url{https://doi.org/10.1103/PhysRevE.64.016131}

\leavevmode\hypertarget{ref-newmanScientificCollaborationNetworks2001}{}%
Newman, M. E. J. (2001b). Scientific collaboration networks. {II}. {Shortest} paths, weighted networks, and centrality. \emph{Physical Review E}, \emph{64}(1), 016132. \url{https://doi.org/10.1103/PhysRevE.64.016132}

\leavevmode\hypertarget{ref-newmanDetectingCommunityStructure2004}{}%
Newman, M. E. J. (2004). Detecting community structure in networks. \emph{The European Physical Journal B - Condensed Matter}, \emph{38}(2), 321--330. \url{https://doi.org/10.1140/epjb/e2004-00124-y}

\leavevmode\hypertarget{ref-rammerKnowledgeProximityFirm2020}{}%
Rammer, C., Kinne, J., \& Blind, K. (2020). Knowledge proximity and firm innovation: {A} microgeographic analysis for {Berlin}. \emph{Urban Studies}, \emph{57}(5), 996--1014. \url{https://doi.org/10.1177/0042098018820241}

\leavevmode\hypertarget{ref-reichardtDetectingFuzzyCommunity2004}{}%
Reichardt, J., \& Bornholdt, S. (2004). Detecting fuzzy community structures in complex networks with a {Potts} model. \emph{Physical Review Letters}, \emph{93}(21), 218701. \url{https://doi.org/10.1103/PhysRevLett.93.218701}

\leavevmode\hypertarget{ref-rossettiCDLIBPythonLibrary2019}{}%
Rossetti, G., Milli, L., \& Cazabet, R. (2019). {CDLIB}: A python library to extract, compare and evaluate communities from complex networks. \emph{Applied Network Science}, \emph{4}(1), 52. \url{https://doi.org/10.1007/s41109-019-0165-9}

\leavevmode\hypertarget{ref-shrumStructuresScientificCollaboration2007}{}%
Shrum, W., Genuth, J., Carlson, W. B., Chompalov, I., \& Bijker, W. E. (2007). \emph{Structures of {Scientific Collaboration}}. {MIT Press}.

\leavevmode\hypertarget{ref-stahlschmidtPerformanceStructuresGerman2019}{}%
Stahlschmidt, S., Stephen, D., \& Hinze, S. (2019). \emph{Performance and {Structures} of the {German Science System}} (p. 91). {Studien zum deutschen Innovationssystem}.

\leavevmode\hypertarget{ref-stephenPerformanceStructuresGerman2020}{}%
Stephen, D., Stahlschmidt, S., \& Hinze, S. (2020). \emph{Performance and {Structures} of the {German Science System} 2020}. {Studien zum deutschen Innovationssystem}.

\leavevmode\hypertarget{ref-taguchiBiMLPACommunityDetection2020}{}%
Taguchi, H., Murata, T., \& Liu, X. (2020). {BiMLPA}: {Community Detection} in {Bipartite Networks} by {Multi}-{Label Propagation}. In N. Masuda, K.-I. Goh, T. Jia, J. Yamanoi, \& H. Sayama (Eds.), \emph{Proceedings of {NetSci}-{X} 2020: {Sixth International Winter School} and {Conference} on {Network Science}} (pp. 17--31). {Cham}: {Springer International Publishing}. \url{https://doi.org/10.1007/978-3-030-38965-9_2}

\leavevmode\hypertarget{ref-traagNarrowScopeResolutionlimitfree2011}{}%
Traag, V. A., Van Dooren, P., \& Nesterov, Y. (2011). Narrow scope for resolution-limit-free community detection. \emph{Physical Review E}, \emph{84}(1), 016114. \url{https://doi.org/10.1103/PhysRevE.84.016114}

\leavevmode\hypertarget{ref-traagLouvainLeidenGuaranteeing2019}{}%
Traag, V. A., Waltman, L., \& van Eck, N. J. (2019). From {Louvain} to {Leiden}: Guaranteeing well-connected communities. \emph{Scientific Reports}, \emph{9}(1), 5233. \url{https://doi.org/10.1038/s41598-019-41695-z}

\leavevmode\hypertarget{ref-tydecksSpatialTopicalImbalances2018}{}%
Tydecks, L., Jeschke, J. M., Wolf, M., Singer, G., \& Tockner, K. (2018). Spatial and topical imbalances in biodiversity research. \emph{PLOS ONE}, \emph{13}(7), e0199327. \url{https://doi.org/10.1371/journal.pone.0199327}

\end{CSLReferences}

\end{document}